\documentstyle[12pt,aaspp4,psfig]{article}

\slugcomment{to appear in Astronomical Journal (Jan 1999)}

\begin{document}

\title{\bf Starburst or Seyfert? Using near-infrared spectroscopy to measure
the activity in composite galaxies}

\author{Tanya~L.~Hill}
\affil{School of Physics, University of Sydney}
\authoraddr{School of Physics, University of Sydney, NSW, 2006, Australia}
\author{Charlene A. Heisler}
\affil{Mount Stromlo and Siding Spring Observatories}
\authoraddr{Mount Stromlo and Siding Spring Observatories, Institute of Advanced Studies, The Australian National University, Private Bag,\\ Weston Creek
Post Office, Canberra, ACT, 2611, Australia\\}
\author{Ralph Sutherland}
\affil{Australian National University Astrophysical Theory Centre and 
Mount Stromlo and Siding Spring Observatories}
\authoraddr{Mount Stromlo and Siding Spring Observatories, Institute of Advanced Studies, The Australian National University, Private Bag,\\ Weston Creek
Post Office, Canberra, ACT, 2611, Australia\\}
\and
\author{Richard W. Hunstead}
\affil{School of Physics, University of Sydney}
\authoraddr{School of Physics, University of Sydney, NSW, 2006, Australia}

\begin{abstract}

We present near-infrared spectra for a sample of galaxies with ambiguous 
optical emission line ratios. These galaxies fall between starbursts and 
Seyferts in the usual optical diagnostic diagrams. We find a similar 
result with the near-infrared emission line ratios, which suggests that the 
galaxies are composite, containing both a starburst and AGN component. 
Furthermore, CO absorption, produced in late-type stars, is detected 
within the sample, although at a weaker level than is typical for 
starburst galaxies. We conclude that the CO feature is being diluted by 
a contribution from an AGN, thereby confirming the composite nature 
of these galaxies.

\end{abstract}

\keywords{galaxies: active --- galaxies: Seyfert --- galaxies: starburst}

\section{Introduction}

The mechanisms by which emission lines are excited in starburst and 
Seyfert galaxies are not fully understood. In simplistic terms, the
gas in starburst galaxies is photoionised by young, hot OB stars,
whereas in Seyfert galaxies, a class of Active Galactic Nuclei 
(AGN), the energy is believed to be derived from photoionisation of 
accreting material around a supermassive black hole and the ionising 
spectrum takes the form of a powerlaw continuum.

There is current debate,
however, over the contribution of star formation to the
energy dynamics of Seyfert galaxies. At one extreme is the starburst
model for AGN (Terlevich {\it et~al.}\ 1992), which is claimed to be able
to reproduce the properties of an AGN using compact supernova remnants (SNRs)
alone, eliminating the need for a black hole entirely. Furthermore, recent 
findings (Boyle \& Terlevich 1998) have suggested that star formation plays 
an important role in QSOs (more luminous AGNs), due to the striking 
similarity between the evolution of the QSO luminosity density and 
galaxy star formation rate.

It is well established that there are some galaxies in which 
both star formation and an AGN contribute
to the observed emission, and their importance for
understanding galaxy energetics is only just being realised.
For example the Seyfert 2 galaxy, NGC 1068, has an extended 
luminous star forming region located $\approx$ 3 kpc from the central 
AGN (Telesco \& Decher 1988), which provides half of the total 
luminosity of NGC 1068 (Lester {\it et~al.}\ 1987; Telesco {\it et~al.}\ 1984). 
Recent observations with the MPE 3D near-infrared (NIR) imaging 
spectrometer have further revealed a stellar core $\approx$~50~pc in size
that provides 7\% of the nuclear bolometric luminosity of NGC 1068
(Thatte {\it et~al.}\ 1997). Likewise, Genzel {\it et~al.}\ (1995) 
have uncovered a circumnuclear ring of 
star formation in the Seyfert 1 galaxy NGC 7469. In this case,
almost two-thirds of the total luminosity of NGC 7469
arises from starburst activity. Furthermore, a compact starburst was
found in Mrk 477, an extremely luminous Seyfert 2 galaxy with a 
hidden Seyfert 1 component. The uncovered starburst is believed
to have a bolometric luminosity comparable to the Seyfert component 
(Heckman {\it et~al.}\ 1997). 

Composite galaxies, i.e., those with both a starburst and Seyfert component,
have implications for evolutionary scenarios in which one type of activity 
may initiate the other. For example, Norman \& Scoville 
(1988) showed theoretically that a black hole can be formed through
the evolution of a massive starburst at the centre of a galaxy.
In the study of NGC 7469, Genzel {\it et~al.}\ (1995) interpret 
a blue-shifted ridge between the nucleus and star formation ring as
infalling gas providing fuel for the AGN. Alternatively, an AGN could 
disturb the gas in a galaxy in such a way as to trigger
star formation, a mechanism proposed for the radio galaxies 3C 285 
(van  Breugel \& Dey 1993) and Minkowski's Object (van Breugel {\it et~al.}\ 1985).

We have undertaken a search of the literature for composite 
galaxies based on optical spectroscopy of their nuclei
as outlined in Section~\ref{sample}. Further in Section~\ref{sample},
we present photoionisation modelling of the optical emission
line ratios that are commonly used as diagnostics. To gain insight into the 
sources of ionisation present within the galaxies we begin
our investigation with NIR spectroscopy. 
Details of the observations and data reduction are given 
in Section~\ref{obsall}. 
The obvious advantage of the NIR is that dust extinction is a factor 
of 10 lower than in the optical, thereby making it possible to probe 
further into the central dust-obscured regions of galaxies where 
the activity occurs. 
We explore several emission and absorption 
features as possible NIR ionisation diagnostics in Section~\ref{nirresults},
with the help of photoionisation models. Throughout this paper, we 
adopt H$_0$ = 75 km s$^{-1}$ Mpc$^{-1}$ and q$_0$ = 0.5.

\section{Sample Selection}
\label{sample}

\subsection{Optical Emission Line Diagnostics}
\label{sampopt}
Starburst galaxies and narrow line AGNs both exhibit emission lines
with FWHM $<1000$ km~s$^{-1}$. Differences in the degree of ionisation 
are traditionally
believed to be due to the hard ionising spectrum of the AGN compared 
with the UV radiation from stars. It is generally accepted that 
AGNs contain an extended partially ionised zone. The hard X-ray component 
of an AGN produces high energy X-ray and far-UV photons with long mean
free paths that can extend beyond the ionised region to 
create a large zone of partially ionised gas. For starburst galaxies,
on the other hand, the transition between neutral and fully ionised 
gas is sharp; the UV radiation is readily absorbed by neutral 
hydrogen, and hence the partially ionised zone is small.
The result is that the forbidden lines, e.g. [O~III], [N~II], [S~II] 
and [O~I], which are collisionally excited and produced mostly
in partially ionised zones, are stronger in Seyfert galaxies 
than in starbursts. On the other hand, hydrogen recombination lines
are stronger in starburst galaxies due to the H~II regions
surrounding young hot OB stars.

Optical emission line ratios are traditionally used to
determine the photoionising source in narrow emission line galaxies.
Veilleux \& Osterbrock (1987) found empirically that the best 
optical diagnostic ratios are those of [O~III]/H$\beta$ compared 
with [N~II]/H$\alpha$, [S~II]/H$\alpha$ and [O~I]/H$\alpha$. These line ratios
are sensitive to the presence of partially ionised gas and are
comprised of easily observable lines spanning a relatively
narrow wavelength range, thereby minimising reddening uncertainties. 

The diagnostic diagrams segregate the galaxies in the manner 
predicted by photoionisation models which use OB stars as the ionising 
source for starburst galaxies and a powerlaw input spectrum for 
AGN (Veilleux \& Osterbrock 1987). However, not all galaxies can be easily 
classified using the diagnostic diagrams (eg. Ashby {\it et~al.}\ 1995). In addition, 
photoionisation may not be the only mechanism operating to produce the 
narrow emission line spectra. The importance of shock excitation over 
photoionisation has been evaluated within the optical regime by 
Dopita \& Sutherland (1996), who show that the spectral characteristics 
of many Seyfert galaxies can be produced by shock excitation alone. 
Shocks can have both a stellar and non-stellar origin, and are believed 
to be associated with supernovae in starburst galaxies and radio jets 
in AGN. Thus, it is important to study in detail those galaxies with 
ambiguous optical line ratio diagnostics in an effort to understand 
the mechanisms involved in producing the observed emission lines. If 
both star formation and an AGN are powering such galaxies, these galaxies
may prove to be important for understanding the relationship, if any, 
between the two energy sources.

\subsection{Sample Selection Criteria}
A search of the literature was undertaken for large optical 
spectroscopic surveys from which to select composite galaxy
candidates, that is, galaxies that potentially contain both intense
star formation and an AGN. We expect that such galaxies would have ambiguous
optical line ratios (intermediate to the starburst and AGN classes)
and therefore used the optical diagnostic
diagrams to choose galaxies that {\em (1)} fell within
$\pm 0.15$dex of the boundary line determined by Veilleux \& Osterbrock
(1987) separating starburst galaxies and AGN 
on all three diagnostic plots, or {\em (2)} fell within 
the domain of starburst galaxies in one diagnostic diagram and AGN
in another. The sample was constrained by a redshift upper limit of $z = 0.035$
and declinations south of +24$^{\circ}$. From the optical spectroscopic 
surveys of Veilleux \& Osterbrock (1987), van den Brock {\it et~al.}\ (1991), 
Ashby {\it et~al.}\ (1995) and Veilleux {\it et~al.}\ (1995), we identified 34
galaxies in which 
the nuclear spectra have ambiguous optical line diagnostics 
satisfying the above criteria. NIR spectroscopy has been obtained for 12
of these galaxies and for a further five galaxies that were observed 
for comparison, comprising one starburst and four AGNs.

\subsection{Modelling the Optical Diagnostics}
\label{model}

While the segregation of starbursts and Seyferts within the optical line 
ratio diagnostic diagrams was discovered observationally, photoionisation 
models (Evans \& Dopita 1985), using hot stars and a powerlaw input 
spectrum to model starbursts and AGN respectively, confirm the observational
finding. However, in the time since these models were calculated, knowledge 
of atomic structure and metal abundances has improved. Therefore, we use 
the latest photoionisation code MAPPINGS~II (see Sutherland \& Dopita 1993) 
to examine the dependence of the optical diagnostic line ratios on 
metallicity (Model A), stellar temperature and hardness of the powerlaw 
spectrum (Model B) and hydrogen density (Model C), according to the parameters
outlined in Table~1. MAPPINGS~II has also been newly expanded so that it 
now includes modelling of NIR emission lines (see Section~\ref{nirresults}). 

The stellar atmosphere models used in MAPPINGS~II are from Hummer 
\& Mihalas (1970) and the AGN models are created from a powerlaw
ionising spectrum of the form $f_\nu \sim \nu^{\alpha}$. 
The ionisation parameter, $U$, is defined by 
$U = (n_H \epsilon^2 l_C)^{1/3}$,
where $n_H$ is the total number density of hydrogen atoms and ions,
$l_C$ is the number of hydrogen-ionising photons emitted
per unit time from the central source, and $\epsilon$ is the volume
filling factor (unity in our models). This parameter 
was varied from $\log U=-3$~to~0
for starburst models and $\log U=-4$~to~$-1$ for AGN models.
Spherical geometry and isochoric conditions were assumed throughout.

The calculations were terminated when the fraction of neutral
hydrogen reached 95\% for both the starburst and AGN powerlaw models,
thus permitting a uniform outer boundary condition.
While a starburst model could be terminated at a higher neutral
fraction, the partially ionised AGN models have numerical
difficulties in achieving complete neutrality, since the
thermal balance becomes due solely to X-ray ionisation and electron 
cascade heating. At a 95\% neutral fraction, the overwhelming majority
of optical and NIR emission from the models has occurred, 
so no significant error is introduced.
The AGN models may not be accurate for far infrared fine structure lines, 
but these are not considered in this paper.

\subsubsection{Model Limitations}
\label{limit}

A simplifying geometric assumption has been built into the models, namely that
photoionisation occurs within a single homogeneous sphere. As a result,
the partially ionised region produced by the models exists as a
very thin, outer shell. In reality, the interstellar medium (ISM) is 
not homogeneous and a thin shell of partially ionised gas will form around 
each density clump found within an ionisation region. This creates
a larger effective emission volume for the [S~II] and [O~I] lines, 
that are formed predominantly in partially ionised regions.
As the ionisation parameter ($U$) is increased within the models,
corresponding to stronger [O~III] emission, the partially ionised region
decreases more sharply than expected.
This is because a non-homogeneous medium will contain clumps of varying 
density, so that even as the ionisation parameter increases the total 
volume of partially ionised gas can remain approximately 
constant. This effect causes the models to slightly underestimate
emission from the partially ionised region, particularly [S~II]
emission. It is most pronounced in starburst models, where
the partially ionised region is necessarily small due to the relatively
soft radiation from stellar sources, compared with the harder ionising
spectrum of a powerlaw model.

Modelling of [O~I] is further limited by the complexity of the 
[O~I] emission process. The collision strengths for transitions of the 
neutral [O~I] species are a strong function of temperature. Further,
collisions by protons and neutral hydrogen atoms can make important 
contributions to [O~I] emission, especially in partially ionised zones that 
occur in AGN. However, the rates for proton and neutral hydrogen 
collisions are not very well known, making accurate predictions 
difficult for [O~I] in AGN (see Williams \& Livio 1995 for reviews 
and discussions on atomic data for emission lines). In starburst 
models, where the [O~I] is confined to a very thin layer in the outer 
radii, the [O~I], and to a lesser extent [S~II], emission is sensitive 
to the chosen boundary conditions.

On the other hand, the [O~III]/H$\beta$ and [N~II]/H$\alpha$ ratios are 
comparatively more reliable, from the 
atomic data and modelling point of view. MAPPINGS II uses a six level 
atomic model for O$^{++}$ and N$^+$. Electron collisions alone 
are the primary excitation mode for the forbidden lines and the 
temperature dependence of the 
collision strengths of these ions is relatively weak and well 
determined. The caveat, however, is that the [N~II]/H$\alpha$ ratio is 
difficult to measure at low spectral resolution due to the blending 
of the [N~II] and H$\alpha$ emission lines.

\subsubsection{Model Analysis}

\begin{figure} [!h]
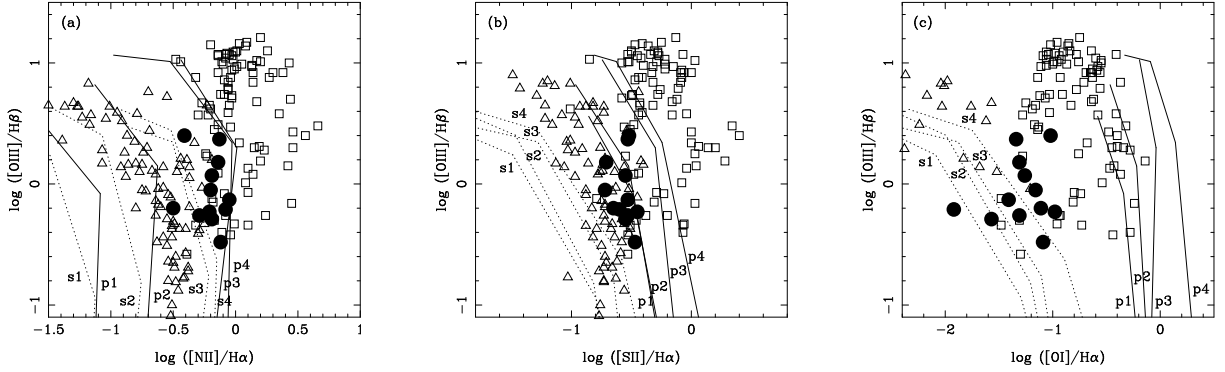

\centerline{\psfig{file=fig1a.ps,bbllx=87pt,bblly=36pt,bburx=585pt,bbury=509pt,height=5cm,angle=270,clip=}
\hspace{7mm}
\psfig{file=fig1b.ps,bbllx=87pt,bblly=36pt,bburx=585pt,bbury=509pt,height=5cm,angle=270,clip=}
\hspace{7mm}
\psfig{file=fig1c.ps,bbllx=87pt,bblly=36pt,bburx=585pt,bbury=509pt,height=5cm,angle=270,clip=}
}
\caption{The optical diagnostic diagrams: 
(a) [O~III]/H$\beta$ vs [N~II]/H$\alpha$; (b) [O~III]/H$\beta$ vs 
[S~II]/H$\alpha$;
(c) [O~III]/H$\beta$ vs [O~I]/H$\alpha$.
Symbols have the following meaning: $\triangle$ -- starbursts and 
$\Box$ -- AGNs from the literature
(Veilleux \& Osterbrock 1987 and references therein);
$\bullet$ -- composite galaxies from this study. The dotted lines are 
starburst models (s1, s2, s3, s4) and the solid lines are powerlaw
models (p1, p2, p3, p4) from MAPPINGS~II showing the effect of increasing
metallicity according to Model A, defined in Table~1. \label{met}}
\end{figure}

\begin{figure} [!h]
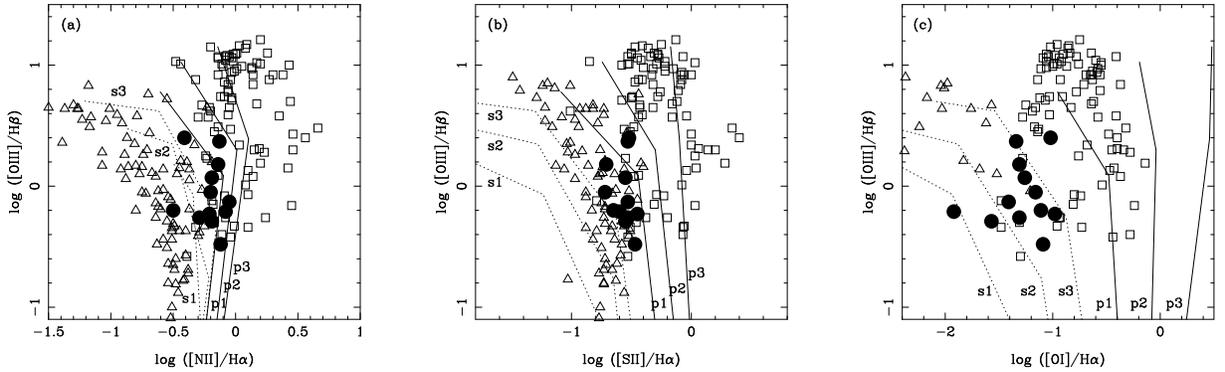

\centerline{\psfig{file=fig2a.ps,bbllx=87pt,bblly=36pt,bburx=585pt,bbury=509pt,height=5cm,angle=270,clip=}
\hspace{7mm}
\psfig{file=fig2b.ps,bbllx=87pt,bblly=36pt,bburx=585pt,bbury=509pt,height=5cm,angle=270,clip=}
\hspace{7mm}
\psfig{file=fig2c.ps,bbllx=87pt,bblly=36pt,bburx=585pt,bbury=509pt,height=5cm,angle=270,clip=}
}
\caption{The optical diagnostic diagrams as shown in 
Figure~\protect\ref{met} with the same symbol meanings.
The starburst (s1, s2, s3) and powerlaw (p1, p2, p3) models
from MAPPINGS~II show the effect of increasing stellar temperature and 
hardness of the powerlaw spectrum respectively, according to Model B,
defined in Table~1. \label{var}}
\end{figure}

\begin{figure} [!h]
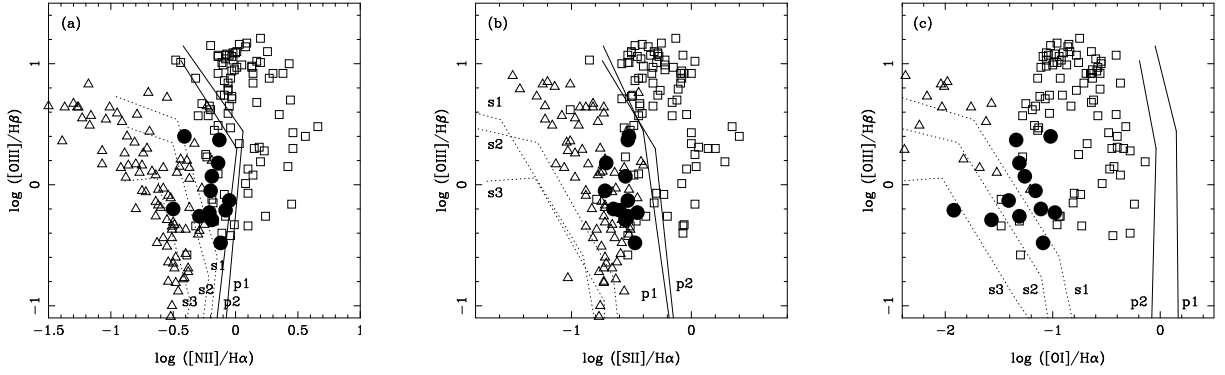

\centerline{\psfig{file=fig3a.ps,bbllx=87pt,bblly=36pt,bburx=585pt,bbury=509pt,height=5cm,angle=270,clip=}
\hspace{7mm}
\psfig{file=fig3b.ps,bbllx=87pt,bblly=36pt,bburx=585pt,bbury=509pt,height=5cm,angle=270,clip=}
\hspace{7mm}
\psfig{file=fig3c.ps,bbllx=87pt,bblly=36pt,bburx=585pt,bbury=509pt,height=5cm,angle=270,clip=}
}
\caption{The optical diagnostic diagrams as shown in 
Figure~\protect\ref{met} with the same symbol meanings. 
The starburst (s1, s2, s3) and the powerlaw (p1, p2) models
from MAPPINGS~II show the effect of decreasing hydrogen density according 
to Model C, defined in Table~1. \label{hden}}
\end{figure}

The results for 
the three different models, defined in Table~1, are shown 
in Figures~\ref{met} -- \ref{hden}, with the starburst models denoted by
`s' and the powerlaw models denoted by `p'. The
observational data are taken from Veilleux \& Osterbrock (1987, and 
papers therein) and we also include our sample of composite galaxies. 

Figure~\ref{met} shows the effect of variation in metallicity. Low
and intermediate metallicities are based on the abundances 
of the SMC and the LMC respectively (Russell \& Dopita 1990). 
The abundances for solar metallicity are taken from Anders \& 
Grevesse (1989) and in a further variation, the solar abundances 
were depleted using the IUE data of Shull (1993). 

The interesting result evident from these models occurs within
the diagnostic diagram of [O~III]/H$\beta$ versus [N~II]/H$\alpha$
(Figure~\ref{met}a). 
It can be seen that when the metallicity is below solar the 
powerlaw models (i.e. p1 and p2) fall into the domain of starbursts.
This suggests that mis-classification of narrow emission line galaxies
using the diagnostic diagrams could occur for low metallicity AGN.

Figure~\ref{var} shows the effect of stellar temperature on the starburst 
models and the hardness of the ionising spectrum on the AGN models.
The diagrams suggest that the starburst 
galaxies are best modelled by high temperatures (40~000~K -- 45~000~K),
while the AGNs require a powerlaw index $\alpha = -1.5$~to~$-2.0$.

Figure~\ref{hden} shows the effect of varying the hydrogen density.
The powerlaw model does not converge for $n_H = 10^2$ cm$^{-3}$, and
therefore is not included in the figure. As expected, starburst galaxies 
are best fitted by models with
$n_H = 10^3$~to~$10^4$~cm$^{-3}$, typical of the H~II regions in
the Orion Nebula. A change in hydrogen density has little effect on the 
powerlaw models.

Overall, the starburst and powerlaw models follow the segregation
of the starburst and AGN data, with the composite galaxies consistently
falling between the two model types.

\section{Observations and Reductions}
\label{obsall}

Echelle spectroscopy was obtained over the period 1994 -- 1996 using the 
InfraRed Imaging Spectrograph (IRIS) at the f/36 Cassegrain focus of the 
3.9~m Anglo-Australian
Telescope (AAT). Two echelles were used, the IJ echelle ($0.9 - 1.5$
$\mu$m) and the HK echelle ($1.46 - 2.5$ $\mu$m), each
with spectral resolution $\lambda/\Delta\lambda \approx 
400$. A non-destructive readout method, with read noise around 40~e$^-$
rms, was used, whereby the array is sampled regularly during an integration. 
This method has the advantage that data acquired after the saturation 
level is reached are removed.
The log of the observations is presented in Table~2. Flux
standards of spectral classes A and G were observed at 
zenith distances similar to the target galaxies. Wavelength calibration
was performed using comparison lamps of helium, argon and xenon.

The galaxy spectra were reduced with the STARLINK program FIGARO
using subroutines written specifically for IRIS. The observations 
were performed in pairs, with the galaxy positioned alternately at
each end of the 13\arcsec-long slit. Subtraction of 
these image pairs provides a good first order sky subtraction and any
residual sky is removed when pairs of extracted spectra are averaged together. 
A flat-field image was formed by observing the dome windscreen illuminated 
by a tungsten lamp; a second exposure with the lamp off was then
subtracted to remove thermal radiation from the telescope and
surroundings. The flat-field image was used to remove wavelength dependent 
pixel-to-pixel variations across the array.

Individual bad pixels were removed from the images by linear interpolation.
Straightening of the echelle orders was achieved using standard star
frames, while the comparison lamp spectra were used to correct for non-vertical
positioning of the slit.
Before using the standard stars to flux calibrate the galaxy spectra and
correct for the atmosphere, hydrogen absorption lines intrinsic to the 
standard were removed. Each standard star spectrum was then fitted to a 
blackbody model with a temperature appropriate for its spectral class.

It is important that approximately the same spatial region be examined for
all the galaxies, to provide consistency in the results. We chose,
therefore, to produce a nuclear spectrum by extracting pixels 
over a window corresponding to a linear scale of $\approx$~1~kpc 
at the redshift of each galaxy. For those galaxies near the redshift upper 
limit of the sample ($z = 0.035$), a linear scale of 1~kpc corresponds to 
a 2~pixel extraction box. IRIS has a pixel scale of 0.79\arcsec/pixel
and therefore the extraction box matches the average seeing 
of 1.5\arcsec, obtained during the observations.

\begin{figure} [!h]
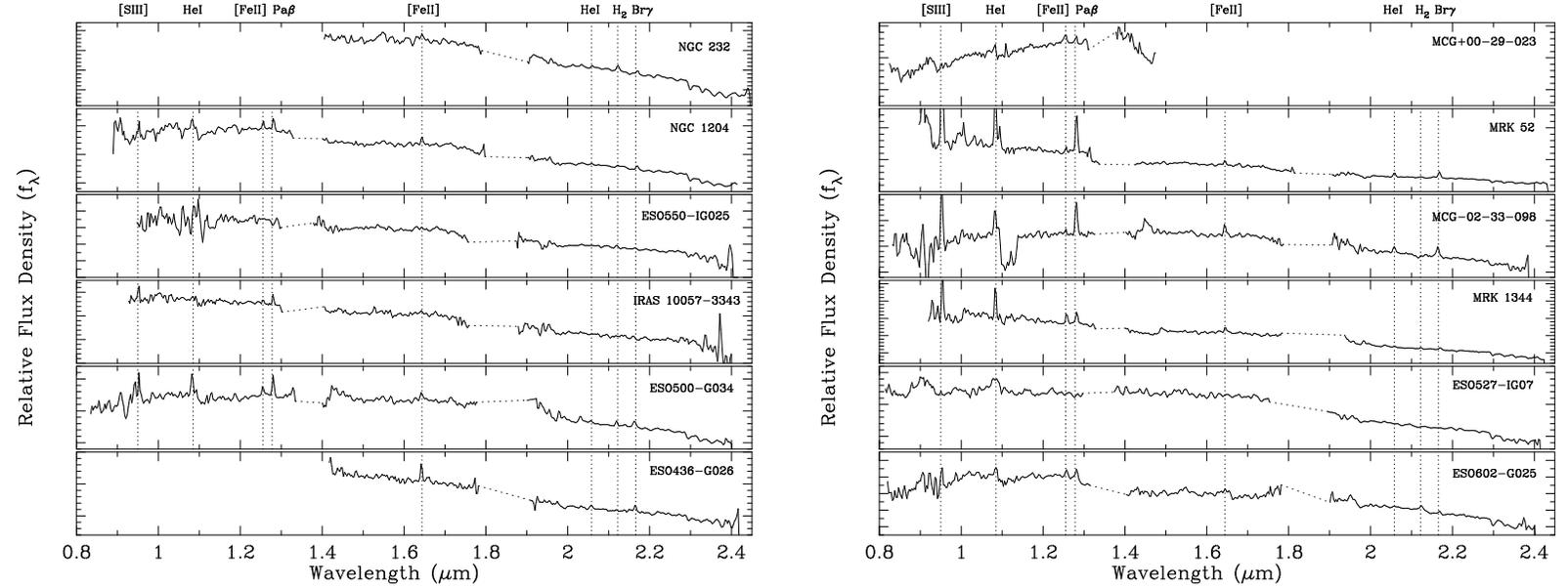

\centerline{\psfig{file=fig4a.ps,bbllx=26pt,bblly=36pt,bburx=585pt,bbury=755pt,height=8cm,angle=270,clip=}
\hspace{4mm}
\psfig{file=fig4b.ps,bbllx=26pt,bblly=36pt,bburx=585pt,bbury=755pt,height=8cm,angle=270,clip=}}
\caption{NIR spectra of the sample of 
composite galaxies. All spectra have been redshift corrected to the
rest frame.
\label{spec1}}
\end{figure}

\begin{figure} [!h]
\centerline{\psfig{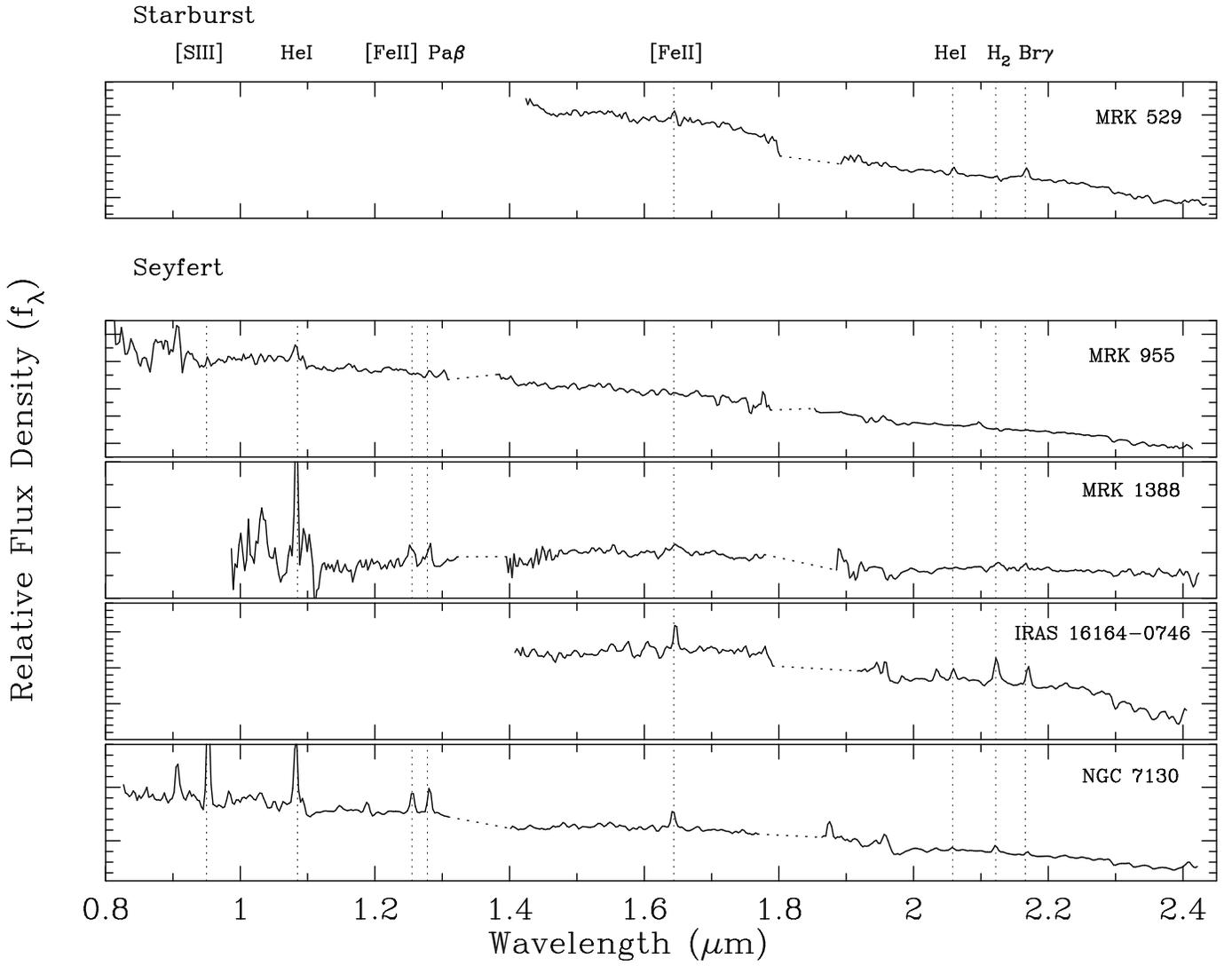}}
\caption{NIR spectra obtained for comparison purposes. 
\label{spec3}}
\end{figure}

The individual orders were combined into one spectrum covering the entire 
IJ and HK bands. This involved bringing all the orders to the same pixel 
scale ($\approx$ 30\AA/pixel). The spectra were shifted to rest wavelength
using published redshifts. Spectra of the sample of composite galaxies 
are shown in Figure~\ref{spec1} and the comparison spectra of starbursts 
and AGN are shown in Figure~\ref{spec3}. The emission line fluxes 
were measured using the IRAF routine SPLOT, which fits a gaussian to 
the line profiles. The relative strengths of the emission lines are
given in Tables~3 and 4 with measurement errors estimated to be
$\approx$10\%.

\begin{figure} [!h]
\centerline{\psfig{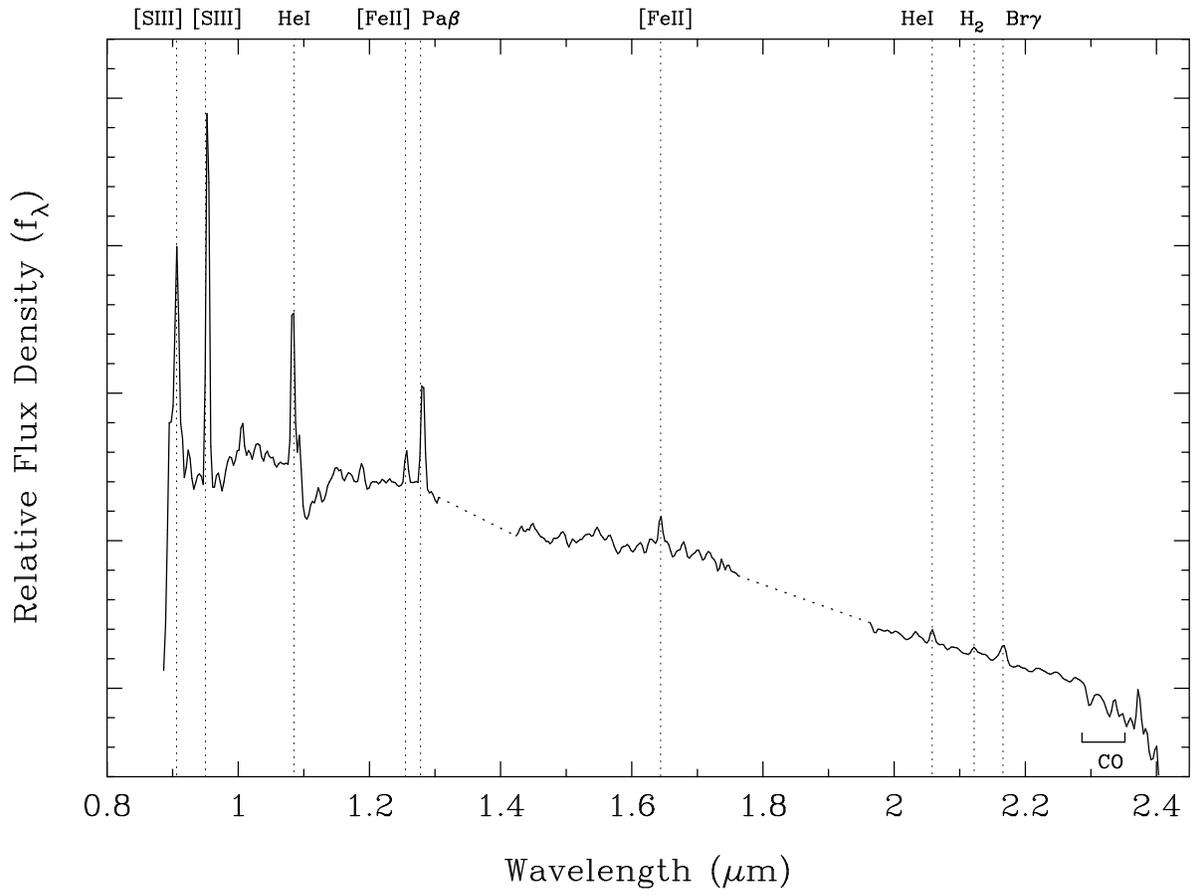}}
\caption{Co-added NIR spectrum formed by averaging together
the spectra of the composite galaxies. \label{comp}}
\end{figure}

The composite galaxies with both IJ and HK data were used to form a
co-added spectrum, which is shown in Figure~\ref{comp}. The individual 
galaxy spectra, all of similar slopes, were first normalised within 
the region 1.5 -- 1.6 $\mu$m and then averaged together. The co-added
spectrum clearly brings out the emission features detected within
the sample.

\section{Results and Discussion}
\label{nirresults}

\subsection{Extinction Corrections}
\label{ext}

Extinction corrections in the optical were calculated following 
Veilleux \& Osterbrock 
(1987), using the Whitford reddening curve parametrized by Miller 
\& Mathews (1972) (see Table~5). The intrinsic H$\alpha$/H$\beta$ 
ratio is usually set at 2.85 for starburst galaxies (assuming Case B 
recombination) and 3.10 for AGN, reflecting enhanced H$\alpha$ emission due 
to collisional excitation. For our sample, where we are unsure of the 
classification, we found that neither value resolved the ambiguous nature 
of the line ratios and chose to adopt H$\alpha$/H$\beta$~=~2.85 for the 
composite galaxies.

Extinction in the NIR can be calculated using the NIR hydrogen 
recombination lines of Pa$\beta$ and Br$\gamma$. However, with IRIS these 
two lines are recorded using different echelles, Pa$\beta$ within IJ and
Br$\gamma$ within HK. A valid comparison requires photometric
conditions during observations in both wavebands.
This occurred for only three galaxies, Mrk 52, ESO~602-G025 and 
NGC 7130 and the values derived for the extinction from the 
NIR lines, E(B--V)$_{\rm{NIR}}$, are 0.0, 1.17 and 1.23, respectively. These
values for E(B--V)$_{\rm{NIR}}$ are consistent with those based on the optical 
line ratios for the three galaxies (see Table~5) implying that the dust is
found within a homogeneous foreground screen, as opposed to it 
being mixed with the line-emitting gas (Puxley \& Brand 1994;
Calzetti {\it et~al.}\ 1996).

\subsection{Emission Features} 

\subsubsection{[Fe~II] Emission}
\label{feii}

In the same manner as the optical forbidden lines, 
[Fe~II]$\lambda\lambda$1.25,1.64
is produced within regions that are partially ionised, suggesting that 
[Fe~II], like [O~I] for example, might be a useful indicator of AGN activity. 
However, [Fe~II] can also be produced via shock excitation. In fact,
strong [Fe~II] emission is found within starburst galaxies,
most likely a result of shock excitation by SNRs
(Mouri {\it et~al.}\ 1993; Forbes \&  Ward 1993; Vanzi \& Rieke 1997),
with a possible contribution from partially ionised 
regions created within cooling zones formed behind the shock front 
in SNRs (Oliva {\it et~al.}\ 1989). SNRs may also contribute to the
[Fe~II] emission in AGN and, in addition, shock excitation can occur
through the interaction of possible jets and outflows from 
the AGN with the surrounding medium. 

There is some indication that [Fe~II] emission may be correlated
with radio emission in starbursts and AGNs (Forbes \& Ward 1993). 
Within starbursts the correlation is readily attributed to shock
excitation from SNRs. Since both starbursts and AGNs follow 
the same correlation it lends support to the idea that SNRs may be 
producing shock excitation within AGN. However, in at least some 
AGNs the shock excitation is more likely associated with jet interactions, 
as confirmed in NGC 1068 (Blietz {\it et~al.}\ 1994) where the [Fe~II] 
emission appears to trace the radio jet.

To further complicate matters, Fe is readily depleted onto dust grains 
and so the destruction of dust grains by shocks can lead to
an enhancement of [Fe~II] as it is released into the gas phase
(Greenhouse {\it et~al.}\ 1991). 

\subsubsection{[Fe~II]/Pa$\beta$ vs [O~I]/H$\alpha$}
\label{oi}

The optical line ratio, [O~I]/H$\alpha$,
is established as a reliable diagnostic for segregating AGN from 
starbursts (Veilleux \& Osterbrock 1987), while the NIR line ratio, 
[Fe~II](1.25$\mu$m)/Pa$\beta$, has also shown promise
as a good diagnostic (Simpson {\it et~al.}\ 1996; Alonso-Herrero 
{\it et~al.}\ 1997) and has a similar advantage of utilising lines close in
wavelength, so reddening effects are minimised. 

In Figure~\ref{niropt} 
we present a plot of [Fe~II]/Pa$\beta$ versus [O~I]/H$\alpha$ for our sample of 
composite galaxies, along with data from the literature (Mouri {\it et~al.}\ 1990, 
1993; Simpson {\it et~al.}\ 1996) and photoionisation model predictions from 
MAPPINGS~II. All the galaxies lie within the region spanned by the Orion
Nebula and SNRs, which may indicate a progression from
pure photoionisation to pure shock excitation (Alonso-Herrero 
{\it et~al.}\ 1997). However, starburst galaxies have been found to lie
along the mixing curve combining H~II regions and
SNRs, whereas Seyfert galaxies do not (Simpson {\it et~al.}\
1996). Our sample galaxies occupy the region between 
starbursts and Seyferts, further supporting their composite nature.

The same photoionisation models used in Section~\ref{model} are included 
in Figure~\ref{niropt}, showing the effect of varying the metallicity,
stellar temperature, powerlaw index and hydrogen density. Our results are
consistent with the CLOUDY models of Alonso-Herrero {\it et~al.}\ (1997),
including the finding that the AGN data are best fitted by powerlaw models 
with low metallicity. However, Alonso-Herrero {\it et~al.}\ (1997) 
find little difference between models with and without grains,
using grain properties derived from the Orion Nebula. Our dust depletion 
model is based on depletion factors in the local ISM (Shull 1993), 
applied to a solar abundance (Anders \& Grevesse 1989). We find 
this model to be too dusty to explain the observed [Fe~II]/Pa$\beta$ ratio,
a reflection of the fact that Fe is very heavily depleted in the local ISM. 

As outlined in Section~\ref{feii}, there is much evidence 
that shock excitation is the dominant mechanism for 
producing [Fe~II] emission. Thus, it is
surprising that the photoionisation models fit the data
so well. Shock models of these ratios using MAPPINGS~II are 
under development (Sutherland 1998).

Two galaxies, Mrk 52 and MCG-02-33-098, have considerably lower 
[Fe~II]/Pa$\beta$ ratios than the rest of the sample. In fact,
Figure~\ref{niropt} provides strong evidence that Mrk 52 
is dominated by star formation, a finding that is further supported
by strong CO absorption and Br$\gamma$ emission (see Section~\ref{cobrg}).
MCG-02-33-098 is an interacting system showing severely disturbed
morphology. Of the two nuclei detected, it is the western nucleus
which has a composite optical spectrum while the eastern nucleus 
has optical line ratios 
consistent with a starburst galaxy. The merger itself may be 
responsible for the unusual line ratios, for example by increasing 
shock excitation by cloud-cloud collisions. However, when 
we examine the optical spectrum for this galaxy (Veilleux {\it et~al.}\ 1995) 
it appears quite noisy and the [O~I]/H$\alpha$ ratio may have 
been overestimated. If so,
MCG-02-33-098 may move closer to the other starburst galaxies
in Figure~\ref{niropt}.

\subsubsection{[Fe~II]/Br$\gamma$ vs H$_2$/Br$\gamma$}
\label{nirdiag}

The usefulness of optical emission line diagnostic diagrams has prompted 
the search for similar diagrams using NIR lines. The lower dust 
extinction in the NIR is an advantage in probing close to the nucleus.
Moorwood and Oliva (1988) proposed [Fe~II](1.644$\mu$m)/Br$\gamma$
versus H$_2$(2.122$\mu$m)/Br$\gamma$ as a possible NIR diagnostic tool
for distinguishing starburst and AGN emission, not least because 
these are the strongest and, therefore, the most easily measured lines.
Their data suggest a segregation of starburst galaxies towards the 
region of low [Fe~II]/Br$\gamma$ and H$_2$/Br$\gamma$
which can readily be explained by the enhanced strength of Br$\gamma$ in 
starburst galaxies compared with AGN (cf. Section~\ref{sampopt}).

For completeness we present the NIR line ratio diagram of 
[Fe~II]/Br$\gamma$ versus H$_2$/Br$\gamma$ in Figure~\ref{nirplot}. Our sample of 
galaxies is plotted together with the original measurements of 
Moorwood \& Oliva (1988), data compiled by Forbes \& Ward (1993)
and blue dwarf galaxies from Vanzi \& Rieke (1997). The composite galaxies, 
like the starbursts and Seyfert~2s, span the range from blue dwarfs 
to Seyfert~1s. Unfortunately, there is no strong separation of 
starbursts and Seyfert~2s, limiting the usefulness of this diagram as
a diagnostic.

\subsubsection{[S~III]/Pa$\beta$ vs [S~II]/H$\alpha$}
\label{siii}

Emission line ratios involving sulfur have been found useful for 
distinguishing between photoionisation and shock excitation (Diaz,
Pagel \& Terlevich 1985; Diaz, Pagel \& Wilson 1985).
If shock excitation is occurring, it is expected that the 
[S~III] line should be weaker than [S~II] since S$^{++}$, in shocked
gas, cools predominantly via UV line emission, corresponding to higher
temperatures, rather than by emission in the NIR (Dopita 1977).

In Figure~\ref{sfig} we present a comparison of the sulfur line 
ratios [S~III]/Pa$\beta$ and [S~II]/H$\alpha$. 
There is a slight tendency for the [S~III]/Pa$\beta$ ratios
of the composite galaxies to favour a starburst origin as
they all lie in the region of log~([S~III]/Pa$\beta$) $< 1$ which
is favoured by starbursts. On the
other hand, the [S~II]/H$\alpha$ line ratio is used in the 
diagnostic diagrams of Veilleux \& Osterbrock (1987) and 
therefore the composite galaxies have been chosen specifically so that 
they lie intermediate between the starbursts and AGN.
SNRs are also 
included in Figure~\ref{sfig}
and fall in the region of low [S~III] emission, as predicted for
shock excitation.

The usual photoionisation models are also shown in Figure~\ref{sfig}.
The powerlaw models are a good fit to the AGN data. However, the starburst
models are a poor fit, with the [S~III]/Pa$\beta$ ratios being too high overall 
to fit the starburst data. As discussed above, shocks can weaken the 
[S~III]/Pa$\beta$ ratio, hinting that shock excitation may be an 
important emission mechanism within the starburst galaxies.

\subsection{Absorption Features}

\subsubsection{CO Indices}
\label{co}

Production of CO occurs in the outer envelopes of late-type stars,
so the CO absorption bandhead longward of 2.3~$\mu$m
appears prominently in the spectra of red giant and supergiant stars. 
Evolutionary modelling (Doyon {\it et~al.}\ 1994) of starburst galaxies 
predicts a sharp increase in CO band strength after approximately
$10^7$ yr, as red supergiants appear in the stellar
population. In principle, therefore, the observed CO absorption is 
a means for constraining the age of a starburst.

It has only been in recent years that NIR spectroscopic detections of CO 
absorption in galaxies have been made and there is no clear consensus
regarding the best method for measuring the strength of the absorption.
The earliest CO detections, from low resolution data, show a
single absorption feature longwards of 2.3~$\mu$m. The original 
spectroscopic CO index (CO$_{\rm{sp}}$) is defined as
\begin{equation}
\rm{CO}_{sp} = -2.5\log_{10}(\langle R_{2.36} \rangle )
\label{coeqn}
\end{equation}
where $\langle R_{2.36}\rangle$ is the average CO depth of the rectified 
spectrum between 2.31 and 2.40~$\mu$m (Doyon {\it et~al.}\ 1994). The 
rectified spectrum is obtained by fitting a powerlaw ($f_\lambda \propto
\lambda^{\beta}$) to featureless regions of the continuum
between 2.00 and 2.29~$\mu$m and extrapolating to longer
wavelengths. CO$_{\rm{sp}}$ was defined in this
manner to be compatible with the photometric CO index (CO$_{\rm{ph}}$)
measured using narrowband filters having effective wavelengths ($\lambda_e$)
and FWHM ($\Delta\lambda$) of $\lambda_e =$2.20, 
$\Delta\lambda =$ 0.11 $\mu$m (continuum) and $\lambda_e =$2.36, 
$\Delta\lambda =$ 0.08 $\mu$m (CO absorption), respectively 
(Frogel {\it et~al.}\ 1978). 
A conversion between CO$_{\rm{sp}}$ and CO$_{\rm{ph}}$ is given by 
Doyon {\it et~al.}\ (1994)
based on the NIR spectroscopic stellar atlas of Kleinmann \& Hall (1986).

The main problem with the calculation of CO$_{\rm{sp}}$ as outlined
above, is the wavelength range of 2.31 -- 2.40 $\mu$m. Many
spectra do not extend to 2.40~$\mu$m due to the increased thermal 
background at longer wavelengths hampering the detection of CO or
because of the limited response of many early NIR arrays.
As a result, in the CO study by Ridgway {\it et~al.}\ (1994), CO$_{\rm{sp}}$
was determined using the same method as Doyon {\it et~al.} (1994)
whereby the average CO depth was measured across a rectified spectrum, 
except in some cases a smaller wavelength region was used, namely, 
2.31 -- 2.37 $\mu$m. The resolution of these early spectra is poor but
comparisons indicated that the results were consistent with 
Doyon {\it et~al.}\ (1994).  
In addition,
Goldader {\it et~al.} (1995) measured CO$_{\rm{sp}}$ in the same manner,
but across an even smaller wavelength region, 2.30 -- 2.34 $\mu$m.

A new method for measuring CO has recently been developed, now that 
increased spectral resolution reveals three distinct absorption 
features: $^{12}$CO~(2,0) at 2.294~$\mu$m, 
$^{12}$CO~(3,1) at 2.323~$\mu$m and $^{13}$CO~(2,0) at 2.345~$\mu$m.
The new CO measurements 
(Oliva {\it et~al.}\ 1995; Puxley {\it et~al.}\ 1997) 
utilize the conventional equivalent width,
\begin{equation}
\rm{EW}(\lambda) = \int_{\lambda_{\rm min}}^{\lambda_{\rm max}} (1 - f'_\lambda)~d\lambda,
\label{coeqn2}
\end{equation}
where $f'_{\lambda}$ is the normalised or rectified spectrum.
Two wavelength regions have been used, the narrow region covering 2.2931 -- 
2.2983~$\mu$m\footnote{Wavelengths quoted in this paper and Puxley 
{\it et~al.}\
(1997) are in vacuum whereas Oliva {\it et~al.}\ (1995) quote the
wavelengths in air.}
and the extended region covering 2.2931 -- 2.3200~$\mu$m.
These regions both include the absorption line $^{12}$CO~(2,0). To compare
with the older measurements, Puxley {\it et~al.}\ (1997) determined 
two transformation equations to convert each equivalent width measurement
(i.e., taken across the narrow region or the extended region) to
the original CO$_{\rm{sp}}$ given in equation~1.

To be consistent with the literature, and to benefit future studies,
we have measured the CO absorption strength using both methods and
various wavelength ranges. Following Puxley {\it et~al.}\ (1997), 
the spectra were normalised by applying 
a powerlaw fit to the continuum ($f_\lambda \propto \lambda^\beta$) tied to 
featureless regions of the spectrum, namely $2.075-2.100$ $\mu$m,
$2.140-2.150$ $\mu$m, $2.235-2.240$ $\mu$m and $2.280-2.290$ $\mu$m.
The values for the CO$_{\rm{sp}}$ index determined directly
from equation~\ref{coeqn} but across various wavelength regions as 
indicated, are given in Table~6. The equivalent width values of CO
measured directly from equation~2, are given in Table~7 for both
the narrow and extended wavelength regions.

It is clear from Table~6 that the CO$_{\rm{sp}}$ index is dependent
on the wavelength region used. We have also tested the transformation
equations derived by Puxley {\it et~al.}\ (1997) to convert from
equivalent width to CO$_{\rm{sp}}$. In Figure~10, we compare 
CO$_{\rm{sp}}$ (calculated), which was calculated from the equivalent
width using the Puxley {\it et~al.}\ (1997) transformation, with
CO$_{\rm{sp}}$ (measured) that has been determined via equation~1
using the wavelength range 2.31 -- 2.40 $\mu$m. This has been
done for both the narrow and extended wavelength regions.
We find that the transformation equation used to convert the
equivalent width across the narrow region to CO$_{\rm{sp}}$
does not work well for our sample of composite galaxies.
The statistical analysis package ASURV (LaValley, Isobe \& Feigelson 1992), 
which implements 
methods for univariate and bivariate problems (Feigelson \& Nelson 
1985; Isobe, Feigelson \& Nelson 1986) showed no significant
correlation between CO$_{\rm{sp}}$ (calculated) and CO$_{\rm{sp}}$
(measured) across the narrow wavelength region. On the other hand,
the transformation equation for the extended
wavelength region works well. The probability that a correlation is not present 
between CO$_{\rm{sp}}$ (calculated) and CO$_{\rm{sp}}$ (measured) 
for the extended region is
0.2\%, and the line of best fit has a gradient of $1.08\pm0.08$. 

In all, we highlight the danger in using different measurement 
methods and then applying transformations to bring the values to 
common ground. Puxley {\it et~al.}\ (1997) suggest that the
equivalent width method should be used in the future to evaluate
the CO absorption and that the optimal wavelength range is
2.2931 -- 2.3200~$\mu$m, that is the extended range. We agree
with this choice as it covers the strongest CO absorption line, yet 
is a large enough region for robust measurements, as shown by
our analysis of CO$_{\rm{sp}}$ (calculated) and CO$_{\rm{sp}}$ (measured). 
The increased resolution of NIR spectra renders the earlier
method of measuring CO$_{\rm{sp}}$ via equation~1 and across large
wavelength regions (2.31 -- 2.40 $\mu$m) obsolete.

\subsubsection{CO Absorption and Br$\gamma$ Emission}
\label{cobrg}

In starburst galaxies, CO absorption and Br$\gamma$ emission 
are both dependent on the age of the stellar population, with CO absorption
arising from older stars and Br$\gamma$ emission resulting from the ionising
UV radiation produced by young hot OB stars. As expected, models show 
a corresponding decrease of Br$\gamma$ emission with increasing
CO absorption as a starburst population ages (Doyon {\it et~al.}\ 1992).

A young starburst, less than 10$^6$ yr old, within a galaxy undergoing its
first burst of star formation would theoretically be expected to exhibit 
strong Br$\gamma$ emission and no CO absorption. This, however, is an extreme 
example and unlikely
to be observed in practice. Studies of low-luminosity dwarf irregular galaxies 
have shown that most, if not all, galaxies have undergone a succession of star
formation episodes (Thuan 1983; Heisler {\it et~al.}\ 1997), making it unlikely 
that we would ever witness a galaxy without a previous generation of stars
contributing to the CO absorption. However, in such galaxies, even with 
an underlying older population, the young massive stars dominate the 
continuum light. For example, blue compact dwarf galaxies have Br$\gamma$
equivalent widths of order 10 nm while little or no
CO absorption is detected (Vanzi \& Rieke 1997).

As the starburst ages and supergiants appear in the population, 
CO absorption increases while Br$\gamma$ emission drops. This is the phase
in which starbursts are generally observed, corresponding to ages between
10$^6$ and 10$^8$~yr, the main-sequence lifetime of OB stars. Beyond 
10$^8$~yr, when all the young OB stars have evolved, the ionising flux
decreases and Br$\gamma$ emission fades, while the CO absorption remains strong. 
At this stage the galaxy has CO absorption characteristic of an elliptical
galaxy,
which can possess CO$_{\rm{sp}}$ values as high as 0.3 (Mobasher \& James 1997).
The strength of CO in elliptical galaxies arises due to metallicity:
giant stars with high metal abundances can have CO absorption 
which is as strong as the more luminous supergiants found in starbursts. 
An old metal rich stellar population, typical of ellipticals, can therefore 
resemble a starburst population in CO absorption.

Our sample of composite galaxies is compared with the data of Oliva 
{\it et~al.}\ (1995) in Figure~\ref{coplot}, where we use the CO equivalent 
width measurements between the wavelengths 2.2931~--~2.2983~$\mu$m.
The life-cycle of a starburst as outlined above, can help to explain the 
distribution of galaxies in this figure. Blue dwarf galaxies 
fall well off Figure~\ref{coplot} to the right, with Br$\gamma$ equivalent 
widths of typically 10~nm and no CO absorption. Starburst galaxies with ages
between 10$^6$ to 10$^8$~yr have consistently high 
CO absorption and a range of Br$\gamma$ 
emission. Older starbursts such as the LMC clusters with ages 
$\geq 10^8$~yr, and elliptical and spiral galaxies, show no Br$\gamma$ and 
strong CO absorption. Few Seyfert galaxies have been investigated for 
CO absorption. The two Seyfert 1 galaxies in the study of Oliva
{\it et~al.}\ (1995) have no detectable Br$\gamma$ and very weak CO absorption.
The Seyfert 2s in the same study have similar CO absorption to ellipticals
and spirals with some Br$\gamma$ emission with equivalent widths of order 
0.1~nm. The weak Br$\gamma$ emission in Seyfert 2s is consistent 
either with emission from the narrow line region or from an old starburst.

With the exception of Mrk 52, the composite galaxies in our sample 
show a range of Br$\gamma$ emission but relatively low CO absorption. 
Mrk 52 has the strongest CO absorption and Br$\gamma$ emission of
all the composite galaxies, supporting the earlier findings
(Section~\ref{oi}) that Mrk 52 is probably dominated by star formation.
The remaining composite galaxies do not have the extremely strong 
Br$\gamma$ emission of young starbursts and the weak CO absorption 
suggests that we are observing older generations of stars that have
already passed through the episode of peak CO production.
However, these are strong optical emission line galaxies 
and so they must contain a nuclear ionising source. If this source is not 
provided by a starburst, it is reasonable to attribute it
to an AGN. To complete the picture, however, an AGN will produce a 
strong continuum that can weaken the CO absorption. Figure~\ref{coagn}
shows the effect of adding an AGN component, modelled by a powerlaw
spectrum of index $\beta = -1.5$, to the spectrum of an M5 giant
star from the atlas of Kleinmann \& Hall (1986). An AGN component
of 50\% results in a drop of $\approx$50\% in the CO absorption. Therefore,
a combination of an AGN plus a starburst population can
explain the range of Br$\gamma$ and CO equivalent widths found for
the composite galaxies and this is also consistent with the NIR and 
optical emission line ratios.

\subsubsection{CO Absorption and Continuum Shape}

When combined with a starburst population, hot dust and/or an 
AGN dilutes the CO equivalent width by adding a strong continuum
component, especially at the red end of the K-band spectrum. 
To examine this effect we combined an AGN component, modelled by
a powerlaw of index $\beta = -1.5$, with the spectrum of a M5 giant
star. The powerlaw and stellar spectra were first normalised to unity
using the average intensity 
between 2.19 -- 2.21 $\mu$m, where the continuum is predominantly
featureless. The spectra were then combined using a powerlaw/stellar
ratio of 0\%, 25\%, 50\% and 75\%, and the results are shown
in Figure~12. Similar modelling showing the dependence of the
CO absorption on extinction and hot dust emission, as well as
the age of the starburst and the upper mass cut-off, can be
found in Schinnerer {\it et~al.}\ (1997).

To determine whether hot dust and/or an AGN is a substantial component
of the sample galaxies we examine the continuum shapes as measured 
by the index $\beta$ against CO$_{\rm{sp}}$ in Figure~\ref{powerlaw}. We 
include 
in Figure~\ref{powerlaw} hot dust models at different temperatures, 
following Goldader {\it et~al.}\ (1997), where a percentage of hot dust is 
added to an M5 giant spectrum (Kleinmann \& Hall 1986). 
The intensity of the hot dust emission $I(\lambda)$ 
is modelled by $I(\lambda) = \lambda^{-1}B(T)$, where 
$B(T)$ is the Planck function (Emerson 1988). 

Four galaxies stand out in Figure~\ref{powerlaw} by having low 
CO$_{\rm{sp}}$ and high $\beta$, consistent with the hot dust models. 
These include one of the AGN
observed by us for comparison, Mrk 1388, and three galaxies from
Ridgway {\it et~al.}\ (1994) which are all classified as Seyferts 
(Sanders {\it et~al.}\ 1988). While none of the composite galaxies show 
such obvious indications of an AGN and/or hot dust, they do, however,
show relatively low values of CO$_{\rm{sp}}$, which from the dust
models may be the result of up to 50\% dust contamination.

The data values in Figure~\ref{powerlaw} have not been corrected 
for extinction, which only affects the continuum shape ($\beta$) and
not CO$_{\rm{sp}}$ (Goldader {\it et~al.}\ 1997; Schinnerer
{\it et~al.}\ 1997). Since NIR extinction cannot be determined directly 
for all the galaxies, either because
the conditions were not photometric or only one of the NIR 
hydrogen recombination lines was observed, we estimate the NIR
extinction from the optical extinction using $A_K = 0.108 A_V$ (Mathis 1990).
The maximum extinction determined for the galaxies ($A_K =$ 0.6), 
is represented by the arrow in Figure~\ref{powerlaw}. Applying such
an extinction to all the composite galaxies would place them
close to or within the region of giants and supergiants, indicating
that the galaxies contain a strong stellar population that is 
affected by dust.

\subsubsection{Other Absorption Features}
\label{nc}

As well as CO absorption, seven of the sample galaxies also show 
Na~I~(2.206 $\mu$m) and/or Ca~I~(2.263 $\mu$m) absorption (see Table~7). These
absorption features are stronger in late-type stars with lower surface 
temperatures, peaking around $T_e = 3 500~$K (Kleinmann \& Hall
1986). Unfortunately, it is not possible to use the Na~I and Ca~I
absorption lines to distinguish between a normal population of
giants and a starburst population of supergiants.
Furthermore, there are added complications due to metallicity, 
with the absorption increasing for higher metal abundances.

Referring back to the optical spectra from the literature,
H$\beta$ absorption is identified in all the composite galaxies.
H$\beta$ absorption is weak in OB stars, peaks in A stars,
before dropping off again with later spectral types. Strong
H$\beta$ absorption therefore generally implies a post-starburst
population. An interesting result is that
AGNs have been found to have a greater mean EW(H$\beta$$_{\rm{abs}}$)
than starburst galaxies (Veilleux {\it et~al.}\ 1995). A statistical 
comparison of the EW(H$\beta$$_{\rm{abs}}$) of
the composite galaxies with the starbursts and AGN using
ASURV, shows that the composite galaxies more closely resemble AGN.
We can reject the hypothesis of the composite galaxies and the
AGN being different at the 95\% confidence level, whereas there 
appears to be no similarity between the composite and starburst samples.

\section{Summary}
\label{summ}

We have defined a sample of galaxies likely to contain both
nuclear star formation and an AGN, based on optical emission
line ratios. NIR spectroscopy of the sample 
has revealed that one galaxy (Mrk~52) is dominated by star formation.
Otherwise, the NIR emission line ratios also
tend to have values intermediate between starbursts and AGN.

The photoionisation code MAPPINGS~II, which models both optical 
and NIR emission lines, was used to confirm the value 
of emission line ratios for classifying emission line galaxies. Within the
diagnostic diagrams, the models agree overall with the segregation 
of starburst and Seyferts seen in the observational data. 
From an investigation of the effect of varying metallicity within the
models we found that low metallicity AGN have the potential to
be mis-classified by the usual optical diagnostics as they fall 
within the regime of the starbursts. Furthermore, 
the [Fe~II]/Pa$\beta$ ratio for AGNs was found
to be best fitted by a powerlaw model of low metallicity.

While shock excitation is often put forward as an important mechanism
for [Fe~II] emission, the photoionisation models provided a good
match to the [Fe~II] data. However, we did find evidence for shock excitation
within the galaxies was found from the sulfur line ratios.

The stellar absorption features detected in the NIR spectra suggest
the presence of an AGN. In particular, when compared with typical starburst 
galaxies the CO absorption is weak within the sample.
We deduce that the sample galaxies are composite in nature:
the starburst produces the CO absorption which is then weakened
by emission from dust heated to temperatures of up to
1000 K by an AGN. 

Further investigations of these galaxies are in progress. As an
additional test for the presence of an AGN in the composite
galaxies, we have undertaken radio continuum imaging and interferometry 
observations, and a paper on these results is in preparation. 
In addition, a program for imaging the galaxies in
the optical and NIR has begun, from which colour maps will be 
formed to deduce the spatial distribution of the dust and star formation
within the galaxies.

\acknowledgments

We would like to thank Mike Dopita, Bahram Mobasher, Jeff Goldader, Peter 
McGregor and Stuart Lumsden for many helpful discussions and ATAC for 
generous allocations of observing time. We also thank the referee
for several suggestions which improved this paper. TLH acknowledges 
the support of 
an Australian Postgraduate Award and RWH acknowledges funding from
the Australian Research Council.
This research has made use of the NASA Astrophysics Data System service
and of the NASA/IPAC Extragalactic Database
which is operated by the Jet Propulsion Laboratory, California Institute of 
Technology, under contract with the National Aeronautics and Space 
Administration.


\begin{figure} [!h]
\centerline{\psfig{file=fig7a.ps,bbllx=87pt,bblly=36pt,bburx=585pt,bbury=509pt,height=5cm,angle=270,clip=}
\hspace{7mm}
\psfig{file=fig7b.ps,bbllx=87pt,bblly=36pt,bburx=585pt,bbury=509pt,height=5cm,angle=270,clip=}
\hspace{7mm}
\psfig{file=fig7c.ps,bbllx=87pt,bblly=36pt,bburx=585pt,bbury=509pt,height=5cm,angle=270,clip=}
}
\caption{The [Fe~II](1.25 $\mu$m)/Pa$\beta$ vs 
[O~I]/H$\alpha$ diagram, repeated to show the three models defined
in Table~1:
(a) shows the effect of varying the metal abundance (Model A);
(b) shows the effect of varying the stellar temperature and powerlaw index
(Model B); and (c) shows the effect of varying the hydrogen density 
(Model C). Symbols
have the following meanings: $\bullet$ -- composite galaxies (optical line
ratios taken from the surveys of Veilleux \& Osterbrock (1987), 
van den Brock {\it et~al.}\ (1991), Ashby {\it et~al.}\ (1995) and Veilleux {\it
 et~al.}\ (1995)); 
$\Box$ -- AGNs (Simpson {\it et~al.}\ 1996);
$\triangle$ -- starbursts (Mouri {\it et~al.}\ 1990); $\times$ -- SNRs
(Mouri {\it et~al.}\ 1993). The Orion Nebula is
also marked on the diagram (Mouri {\it et~al.}\ 1993). The model labels 
are defined in Table~1. \label{niropt}} 
\end{figure}

\begin{figure} [!h]
\centerline{\psfig{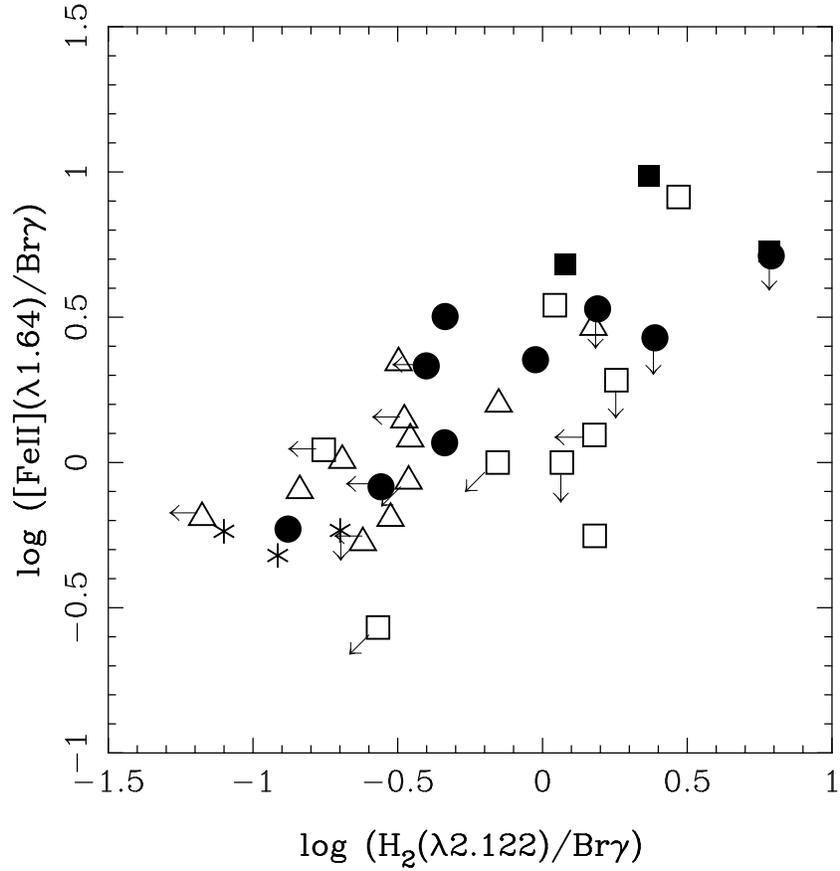}}
\caption{A possible NIR diagnostic diagram of
[Fe~II](1.64 $\mu$m)/Br$\gamma$ vs H$_2$(2.12 $\mu$m)/Br$\gamma$. 
The symbols have
the following meanings: $\bullet$ -- composite galaxies;
\protect\rule[0.5mm]{2mm}{2mm} -- Seyfert 1s; $\Box$ -- Seyfert 2s; 
$\triangle$ -- starbursts (Moorwood \& Oliva 1988; Forbes \& Ward 1993); 
$\ast$ -- blue dwarf galaxies (Vanzi \& Rieke 1997). Limits are 
shown with arrows. \label{nirplot}}
\end{figure}

\begin{figure} [!h]
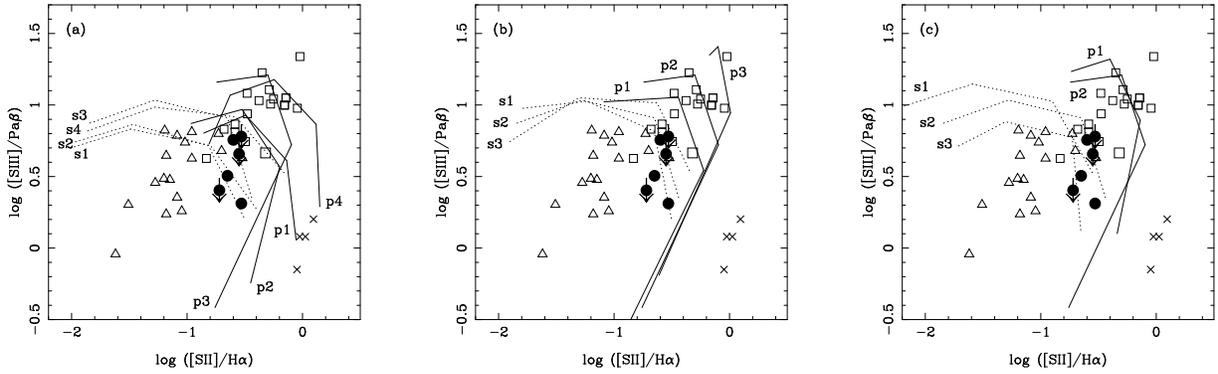

\centerline{\psfig{file=fig9a.ps,bbllx=87pt,bblly=36pt,bburx=585pt,bbury=509pt,height=5cm,angle=270,clip=}
\hspace{7mm}
\psfig{file=fig9b.ps,bbllx=87pt,bblly=36pt,bburx=585pt,bbury=509pt,height=5cm,angle=270,clip=}
\hspace{7mm}
\psfig{file=fig9c.ps,bbllx=87pt,bblly=36pt,bburx=585pt,bbury=509pt,height=5cm,angle=270,clip=}
}
\caption{The [S~III]/Pa$\beta$ vs [S~II]/H$\alpha$ 
diagram, repeated to show the three models defined in Table~1:
(a) shows the effect of varying the metal abundance (Model A);
(b) shows the effect of varying the stellar temperature and powerlaw index
(Model B); and (c) shows the effect of varying the hydrogen density 
(Model C). Symbols have
the following meanings: $\bullet$ -- composite galaxies (optical line
ratios taken from the surveys of Veilleux \& Osterbrock (1987),
van den Brock {\it et~al.}\ (1991), Ashby {\it et~al.}\ (1995) and Veilleux {\it
 et~al.}\ (1995)); 
$\Box$ -- AGNs;
$\triangle$ -- starbursts; $\times$ -- SNRs (Kirhakos \& Phillip
1989; Osterbrock {\it et~al.}\ 1992).  The model labels are defined
in Table~1. \label{sfig}}
\end{figure}
\begin{figure} [!h]
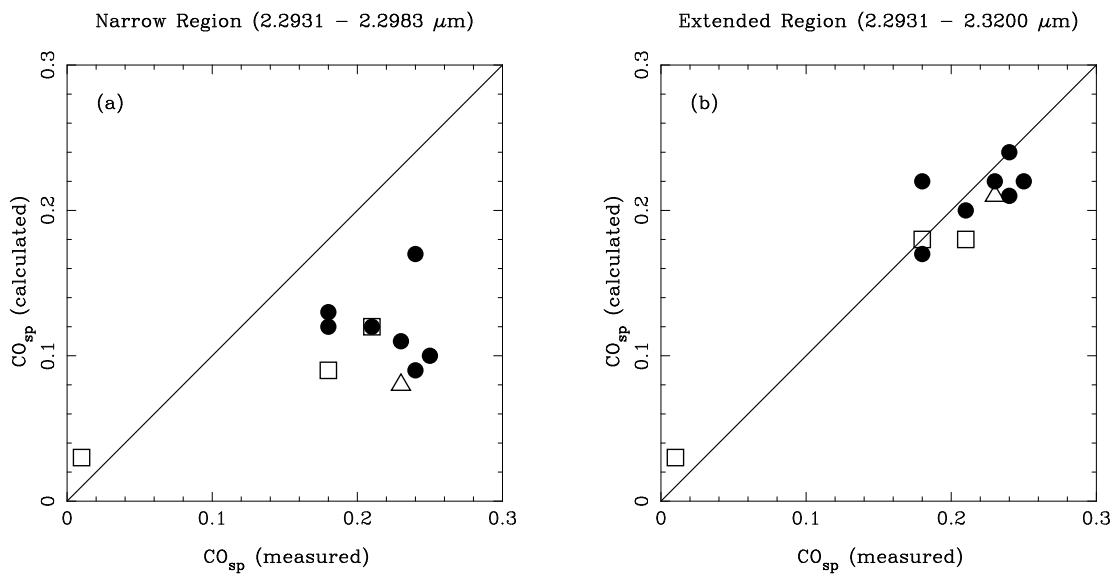

\centerline{\psfig{file=fig10a.ps,bbllx=30pt,bblly=16pt,bburx=605pt,bbury=529pt,height=8cm,angle=270,clip=}
\hspace{5mm}
\psfig{file=fig10b.ps,bbllx=30pt,bblly=16pt,bburx=605pt,bbury=529pt,height=8cm,angle=270,clip=}}
\caption{CO$_{\rm{sp}}$ (measured) was obtained directly 
from equation~1 across a wavelength region of
2.31~--~2.40 $\mu$m. CO$_{\rm{sp}}$ (calculated) was obtained from 
the CO absorption equivalent widths and applying the
transformation equations of Puxley {\it et~al.}\ (1997). 
(a) is for the narrow region (2.2931~--~2.2983 $\mu$m) and 
(b) is for the extended region (2.2931~--~2.3200 $\mu$m).
Symbols have 
the usual meanings: $\bullet$~--~composites; 
$\triangle$~--~starbursts; $\Box$~--~Seyfert 2s. The
straight line defines equality between the two estimates. \label{cor}}
\end{figure}

\begin{figure} [!h]
\centerline{\psfig{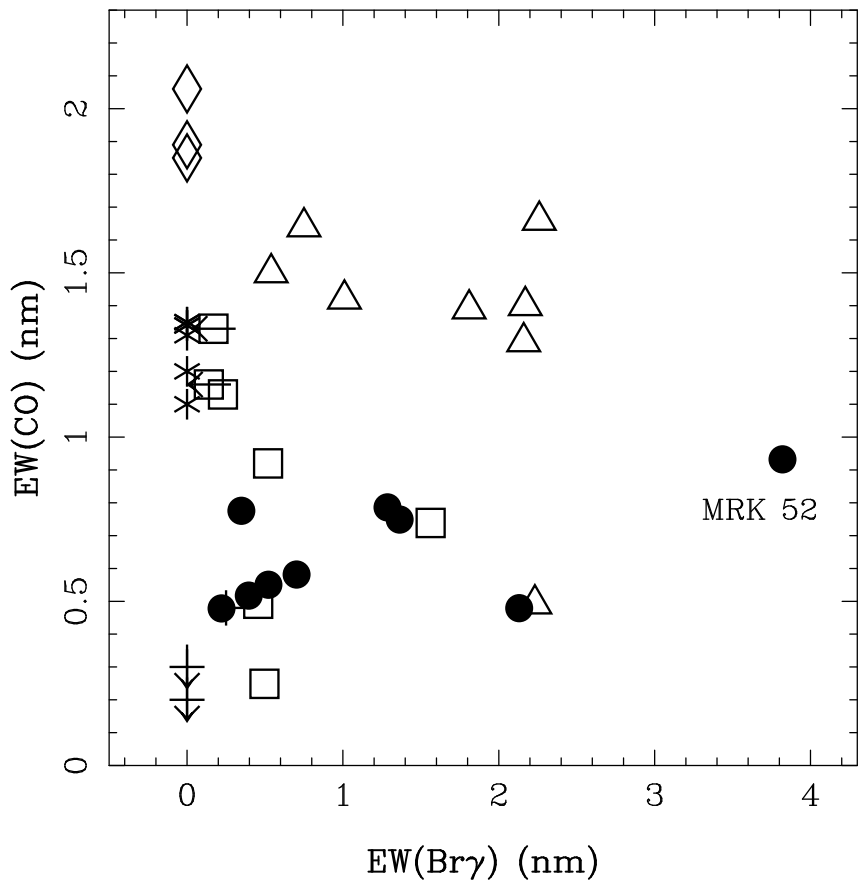}}
\caption{The equivalent width of CO, measured within
the narrow region (2.2931 -- 2.2983~$\mu$m), plotted against the 
equivalent width of Br$\gamma$. While we believe the narrow region is
not the best wavelength range to be used for measuring CO, it
was necessary so that a consistent comparison could
be made between the sample, shown as filled circles, and
data from Oliva {\it et~al.}\ (1995). The symbols have the
following meaning: $\triangle$ -- starbursts; $\Box$ -- Seyfert 2s;
+ -- Seyfert 1s; $\ast$ -- ellipticals and spirals; $\Diamond$ --
LMC clusters with ages $<$8~Myr. Limits are shown with arrows.
Note that blue dwarf galaxies 
fall outside the boundaries of this diagram, having 
EW(Br$\gamma$)$\sim$10~nm and no CO absorption (Vanzi \& Rieke 1997). 
\label{coplot}}
\end{figure}

\begin{figure} [!h]
\centerline{\psfig{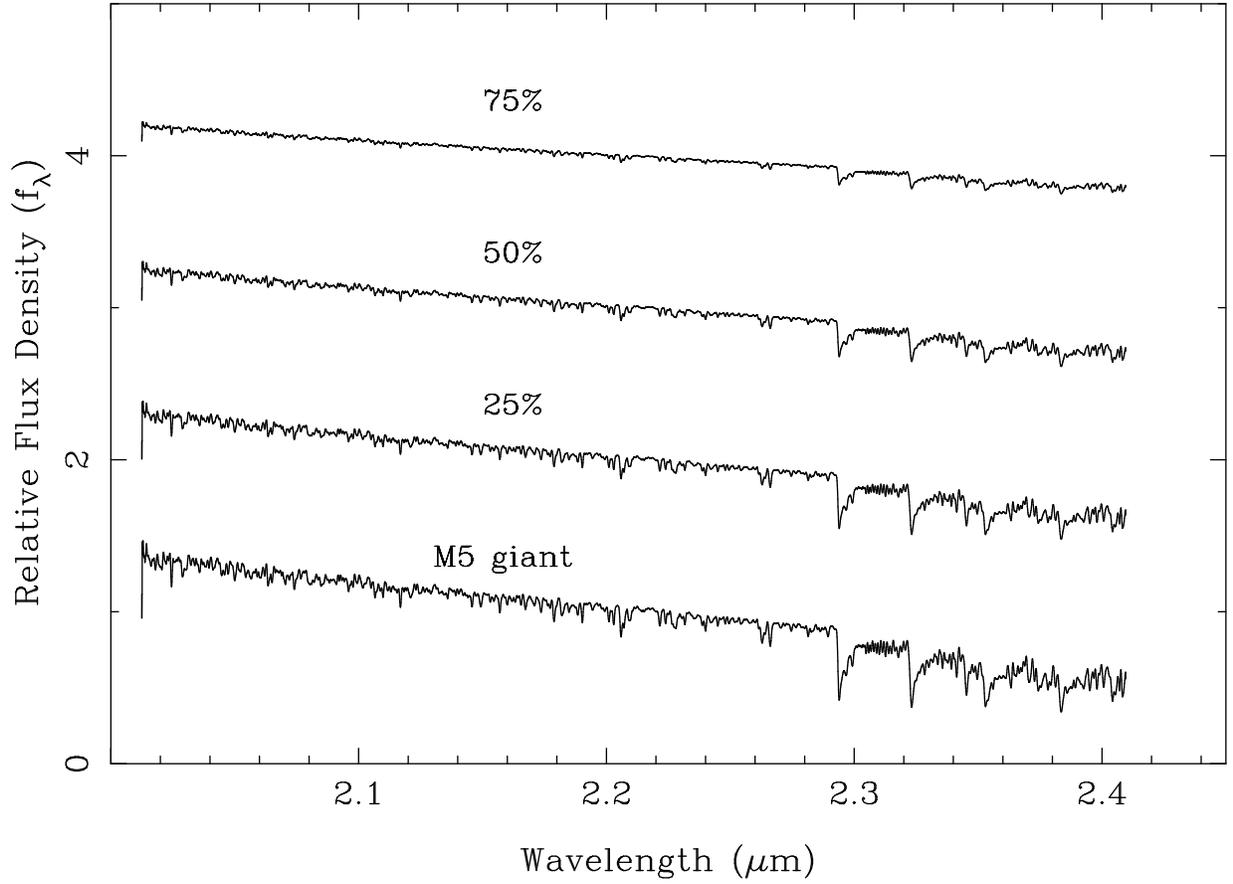}}
\caption[hill.f12.ps]{We examine the effect of adding an
AGN component to the spectrum of a M5 giant star in a ratio
of 25\%, 50\% and 75\% AGN. The AGN
component is modelled by a powerlaw of index $\beta = - 1.5$.
All spectra are shown to the same scale and have been shifted vertically
for easy comparison. \label{coagn}}
\end{figure}

\begin{figure} [!h]
\centerline{\psfig{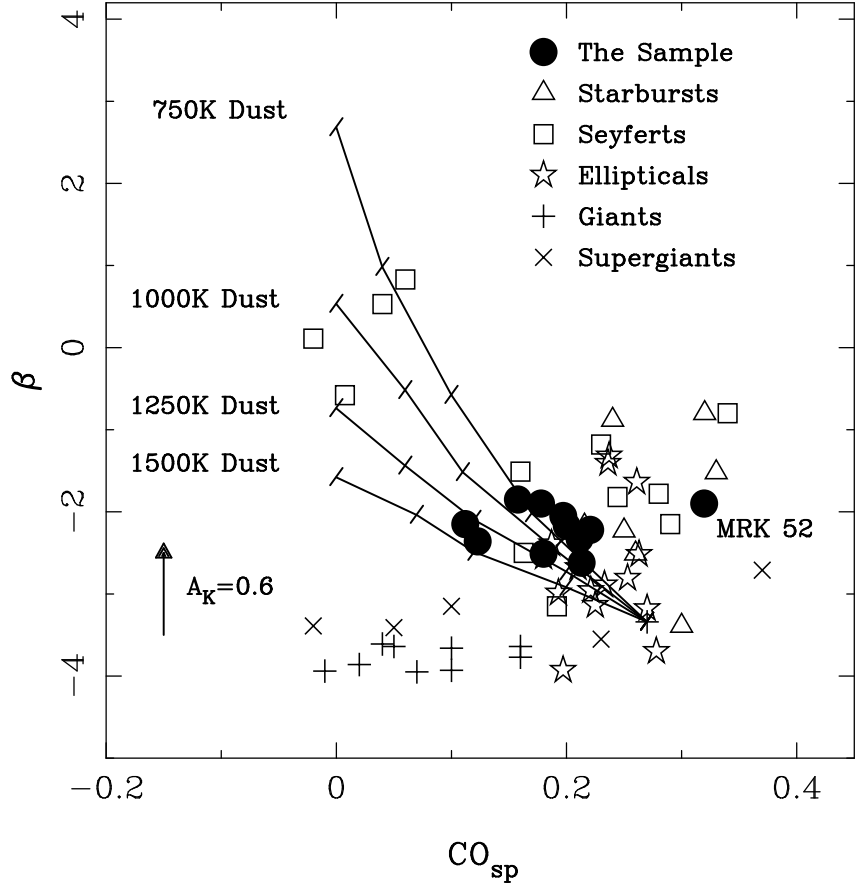}}
\caption{The CO$_{\rm{sp}}$ index has been measured 
according to Ridgway {\it et~al.}\ (1994), following equation~1 and
using a wavelength region of 2.31 -- 2.37 $\mu$m.
While this is no longer the optimal method for measuring CO, it
was necessary for consistent comparison with literature data.
The CO$_{\rm{sp}}$ index is plotted against the index of the
powerlaw fit to the continuum (i.e. $f_\lambda \propto \lambda^\beta$).
With our composite galaxy data (filled circles) we include
starbursts and Seyferts (Ridgway {\it et~al.}\ 1994), ellipticals
(Mobasher \& James 1997) and giants and supergiants from the
stellar atlas of Kleinmann \& Hall (1986). The curves are dust
models at different temperatures, with markings representing a
dust component added to a M5 giant spectrum in the ratio of 25\%, 50\%,
75\% and 100\%. The arrow shows how the addition of extinction moves
a point within the diagram. \label{powerlaw}}
\end{figure}

\clearpage
\newpage

\begin{deluxetable} {lccc} 
\tablewidth{0pt}
\tablenum{1}
\tablecaption{Photoionisation model parameters \label{mod}}
\tablehead{
\colhead{Parameter} &
\colhead{Model A} &
\colhead{Model B} &
\colhead{Model C} \nl
}
\tablecolumns{4}
\startdata
Metallicity & s1, p1: SMC & & \nl 
& s2, p2: LMC & solar & solar \nl
& s3, p3: solar & & \nl
& s4, p4: depleted solar & & \nl \nl
Stellar Temperature (K) & & s1: 38 000 & \nl
& 40 000  & s2: 40 000  & 40 000 \nl
& & s3: 45 000  & \nl \nl
Powerlaw Index ($\alpha$) & & p1: $-$2.0 & \nl
& $-$1.5  & p2: $-$1.5 & $-$1.5 \nl
& & p3: $-$1.0 & \nl \nl
Hydrogen Density (cm$^{-3}$) & & & s1, p1: $10^4$ \nl
& $10^3$ & $10^3$ & s2, p2: $10^3$ \nl
& & & s3: $10^2$ \nl
\enddata
\tablecomments{The starburst photoionisation models, denoted by `s', 
use the stellar atmosphere models of Hummer \& Mihalas (1970). The
powerlaw models, denoted by `p', use a powerlaw ionising 
source of the form $f_\nu \sim \nu^{\alpha}$. 
The metal abundances used are based on
the SMC and LMC (Russell \& Dopita 1990), representing low and
intermediate metallicities respectively, the solar abundances
are from Anders \& Grevesse (1989) and the depletion factors applied
to the solar abundances were taken from Shull (1993).}
\end{deluxetable}


\begin{deluxetable} {lllccccc}  
\footnotesize
\tablenum{2}
\tablewidth{0pt}
\tablecaption{Log of AAT observations \label{obs}}
\tablehead{
\colhead{Name} &
\colhead{RA(2000)} &
\colhead{Dec(2000)} &
\colhead{z}  &
\colhead{Echelle} &
\colhead{Date} &
\colhead{Int. Time}  &
\colhead{Obs.} \nl
\colhead{} & \colhead{} & \colhead{} & \colhead{} & \colhead{} &
\colhead{} & \colhead{(s)} & \colhead{Cond.} 
}
\tablecolumns{8}
\startdata
NGC 232 & 00 42 45.9 & $-$23 33 36 & 0.023 & HK & 15 Aug 95 & 4800 
& N \nl
NGC 1204 & 03 04 40.4 & $-$12 20 28 & 0.015 & HK & 12 Aug 95 & 3600 & P \\ 
& & & & IJ & 13 Aug 95 & 1800 & P \nl
ESO 550-IG025 N & 04 21 19.9 & $-$18 48 39 & 0.032 & HK & 6 Feb 96 & 3600 & N \nl
& & & & IJ & 7 Feb 96 & 900 & N \nl
IRAS 10057-3343 & 10 07 59.1 & $-$33 58 07 & 0.034 & HK & 7 Feb 96 &  3600 & N \nl
& & & & IJ & 7 Feb 96 & 5400 & N \nl
ESO 500-G034 & 10 24 31.4 & $-$23 33 11 & 0.013 & HK & 5 Feb 96 &  3600 & N \nl
& & & & IJ & 5 Feb 96 & 5400 & N \nl
ESO 436-G026 & 10 28 42.7 & $-$31 02 18 & 0.014 & HK & 5 Feb 96 &  3600 & N \nl
MCG+00-29-023 & 11 21 11.7 & $-$02 59 03 & 0.025 & IJ & 7 Feb 96 & 5400 & P \nl
Mrk 52 (NGC 4385) & 12 25 42.6 & +00 34 23 & 0.007 & HK &  16 May 94 &  1920 & 
P \nl
& & & & IJ & 16 May 94 & 3200 & P \nl
MCG-02-33-098 W & 13 02 19.9 & $-$15 46 06 & 0.017 & HK & 6 Feb 96 & 600 & N \nl
& & & & IJ & 6 Feb 96 & 900 & N \nl
Mrk 1344 (NGC 4990) & 13 09 17.2 & $-$05 16 23 & 0.011 & HK & 17 May 94 &  3000 & N \nl
ESO 527-IG07 & 20 04 31.3 & $-$26 25 40  & 0.035& HK & 14 Aug 95 & 4800 & N \nl
& & & & IJ & 14 Aug 95 & 7200 & N \nl
ESO 602-G025 & 22 31 25.3 & $-$19 02 05 & 0.025 & HK & 31 Oct 95 & 3600 & P \nl
& & & & IJ & 1 Nov 95 &  5400 & P \nl
\cutinhead{Starbursts}
Mrk 529 (NGC 7532) & 23 14 22.2 & $-$02 43 39 &  0.010  &  HK & 14 Aug 95 & 2400 & P \nl
\cutinhead{AGN}
Mrk 955  & 00 37 35.8 & +00 16 51& 0.035 & HK & 13 Aug 95 &  3600 & P \nl
& & & & IJ & 13 Aug 95 &  5400 & P \nl
Mrk 1388 & 14 50 37.7 & +22 44 04 & 0.021 & HK & 18 May 94  &  2400 & N \nl
& & & & IJ & 18 May 94 & 2400 & N \nl
IRAS 16164-0746 & 16 19 10.1 & $-$07 53 57 & 0.021 & HK & 12 Aug 95 & 3600 & P \nl
& & & & IJ & 12 Aug 95 &  3600 & P \nl
NGC 7130 (IC 5135) & 21 48 19.3 & $-$34 57 03 & 0.016& HK & 13 Aug 95 &  3600 & P \nl
& & & & IJ &13 Aug 95 &    5400 & P \nl
\enddata
\tablecomments{Observing conditions: N -- Non-photometric; P -- Photometric.}
\end{deluxetable}


\begin{deluxetable} {lcccccc} 
\small
\tablewidth{0pt}
\tablenum{3}
\tablecaption{Relative intensities of emission lines in the IJ spectra \label{ijdata}}
\tablehead{
\colhead{Name} &
\colhead{[S~III]} & \colhead{[S~III]} & 
\colhead{He~I} & \colhead{[Fe~II]} & 
\colhead{Pa$\beta$} & \colhead{Comments}  \nl
& \colhead{$\lambda$0.90} & \colhead{$\lambda$0.95} & 
\colhead{$\lambda$1.085} & \colhead{$\lambda$1.25} & 
\colhead{$\lambda$1.28} & } 
\tablecolumns{7}
\startdata
NGC 232 & \nodata & \nodata & \nodata & \nodata & \nodata & no IJ spectrum \nl
NGC 1204 & \nodata &  0.81 & 1.59 & 0.62 & 1.0 & \nl
ESO 550-IG025 N & \nodata & \nodata & \nodata & \nodata & \nodata & no measurable
 lines \nl
IRAS 10057-3343 & \llap{$<$}4.70 & \llap{$<$}1.33 & \llap{$<$}1.20 & 0.40 & 1.0 & \nl
ESO 500-G034 & 1.09 & 0.96 & 1.35 & 0.54 & 1.0 & \nl
ESO 436-G026 &\nodata & \nodata & \nodata & \nodata & \nodata & no IJ spectrum \nl
MCG+00-29-023 & \llap{$<$}2.60 & 1.95 & 1.33 &  1.97 &  1.0 & \nl
Mrk 52 & 1.56 & 4.15 & 1.57 & 0.09 & 1.0 & \nl
MCG-02-33-098 W & 1.24 &  1.96 & 0.96 & 0.09 & 1.0 & \nl
Mrk 1344 & \nodata & 3.32 & 2.06 & 0.77 & 1.0 & \nl
ESO 527-IG07 &  \nodata & \nodata & \nodata & \nodata & \nodata & no measurable
lines \nl
ESO 602-G025 & \llap{$<$}1.48 & 1.06 & 0.96&  0.62 &  1.0 & \nl 
\cutinhead{Starbursts} 
Mrk 529  & \nodata & \nodata & \nodata & \nodata & \nodata & no IJ spectrum \nl
\cutinhead{AGN} 
Mrk 955 & \nodata & 1.10 & 6.58 & \llap{$<$}1.78 & 1.0 & \nl
Mrk 1388 & \nodata &  \nodata & 5.01 & 1.18 & 1.0 & \nl
IR 16164-0746 & \nodata & \nodata & \nodata & \nodata & \nodata & no IJ spectrum\nl
NGC 7130  & 1.38 & 3.25 & 2.77 & 0.85 & 1.0 & \nl
\enddata
\end{deluxetable}

\begin{deluxetable} {lccccccc} 
\small
\tablewidth{0pt}
\tablenum{4}
\tablecaption{Relative intensities of emission lines in the HK spectra \label{hkdata}}
\tablehead{
\colhead{Name} &
\colhead{H$_2$} & \colhead{He~I} & \colhead{H$_2$} & \colhead{Br$\gamma$} & 
\colhead{H$_2$} & \colhead{H$_2$} & \colhead{Comments} \nl 
& \colhead{$\lambda$2.034} & \colhead{$\lambda$2.058} & \colhead{$\lambda$2.122} & 
\colhead{$\lambda$2.166} & \colhead{$\lambda$2.223} & \colhead{$\lambda$2.248} &} 
\tablecolumns{8}
\startdata
NGC 232 & \llap{$<$}0.90 & 1.05 & 2.45 & 1.0 & 1.02 & 0.21 & \nl
NGC 1204 & \llap{$<$}0.85 & \llap{$<$}0.47 & 0.94 & 1.0 & 0.43 & 0.49 & 20\% error in H$_2$(2.122) \nl
ESO 550-IG025 N & \nodata & \llap{$<$}0.55 & 1.55 & 1.0 & \nodata & \nodata & \nl
IRAS 10057-3343 & \nodata & \nodata & \nodata & \nodata & \nodata & \nodata & no
 measurable lines \nl
ESO 500-G034 & 0.68 & 0.25 & 0.46 & 1.0 & 0.46 & 0.17 & \nl
ESO 436-G026 & \nodata & 0.40 & \llap{$<$}0.40 & 1.0 & \nodata & \nodata & \nl
MCG+00-29-023 &  \nodata & \nodata & \nodata & \nodata & \nodata & \nodata & no
HK spectrum \nl
Mrk 52 & \nodata & 0.45 & 0.13 & 1.0 & \nodata & \nodata & \nl
MCG-02-33-098 W & \nodata & 0.41 & \llap{$<$}0.28 & 1.0 & \nodata & \nodata & \nl
Mrk 1344 & 1.21 & 0.52 & 0.46 & 1.0 & 0.53 & \llap{$<$}0.11 &  \nl
ESO 527-IG07 & \nodata & \nodata & \nodata & \nodata & \nodata & \nodata & no measurable lines \nl
ESO 602-G025 & \nodata & \llap{$<$}1.71 & 6.17 & 1.0 & \nodata & \nodata & \nl
\cutinhead{Starbursts}
Mrk 529  & \nodata & 0.25 & \llap{$<$}0.24 & 1.0 & \nodata & \nodata & \nl
\cutinhead{AGN}
Mrk 955 &   \nodata & \nodata & \nodata & \nodata & \nodata & \nodata & no measurable lines \nl
Mrk 1388 & \nodata & \llap{$<$}0.35 & 1.80 & 1.0 & \nodata & \nodata & \nl
IRAS 16164-0746 & 0.20 & 0.28 & 1.51 & 1.0 & 0.50 & 0.33 & \nl
NGC 7130  & \nodata & 0.59 & 2.95 & 1.0 & 1.12 & 0.54 & \nl
\enddata
\end{deluxetable}

\begin{deluxetable} {lcccc} 
\small
\tablewidth{0pt}
\tablenum{5}
\tablecaption{Optical extinction data taken from the literature
\label{exttab}}
\tablehead{
\colhead{Name} &
\colhead{H$\alpha$/H$\beta$} &
\colhead{E(B$-$V)} &
\colhead{FWHM} &
\colhead{Ref.} \nl
\colhead{} & \colhead{} & \colhead{} & \colhead{(km s$^{-1}$)} & 
\colhead{} 
}
\tablecolumns{5}
\startdata
NGC 232 & 10.5 & 1.32 & 120 & 1 \nl
NGC 1204 & 17.4 & 1.83 & \nodata & 1 \nl
ESO 550-IG025 N & 9.12 & 1.18 & 830  & 1 \nl
IRAS 10057-3343 & 17.78 & 1.86 & 300 & 2 \nl
ESO 500-G034 & 18.20 & 1.88 & 300 & 2 \nl
ESO 436-G026 & 10.72 & 1.34  & 300 & 2 \nl
MCG+00-29-023 & 12.3 & 1.48 & 660 & 1  \nl
Mrk 52 & 2.41 & 0.0 & \llap{$<$}260 & 3 \nl
MCG-02-33-098 W & 12.3 & 1.48 & 1910 & 1 \nl
Mrk 1344 & 10.23 & 1.30 & \nodata &  4  \nl
ESO 527-IG07 & 9.55 & 1.22 & 200 & 2 \nl
ESO 602-G025 & 10.7 & 1.34 & 800 & 1 \nl
\cutinhead{Starbursts}
Mrk 529  & 5.50 & 0.67 & \nodata & 4  \nl
\cutinhead{AGN}
Mrk 955 & 8.13 & 0.98  & \nodata & 4 \nl
Mrk 1388 & 4.57 & 0.39 & \nodata & 4 \nl
IR 16164-0746 & 16.6 & 1.70 & \nodata & 1 \nl
NGC 7130  & 11.22 & 1.30 & 600 & 2 \nl
\enddata
\tablerefs{
(1)~Veilleux {\it et~al.} 1995; (2)~van den Broek {\it et~al.} 1991; 
(3)~Durrant \& Tarrab 1988; (4)~Veilleux \& Osterbrock 1987.}
\end{deluxetable}


\begin{deluxetable} {lccccccc}  
\small
\tablewidth{0pt}
\tablenum{6}
\tablecaption{The spectroscopic CO indices and related measurements \label{co}}
\tablehead{
\colhead{Name} & \colhead{CO$_{\rm{sp}}$} & 
\colhead{CO$_{\rm{sp}}$} & \colhead{CO$_{\rm{sp}}$} 
& \colhead{CO$_{\rm{sp}}$} & 
\colhead{CO$_{\rm{sp}}$} &  \colhead{EW(Br$\gamma$$_{\rm{em}}$)}
& \colhead{$\beta$} \nl
& \colhead{2.31 -- 2.37} & \colhead{2.31 -- 2.40} & \colhead{extended} 
& \colhead{narrow} & \colhead{2.30 -- 2.34} & \colhead{nm} & \colhead{} \nl
\colhead{(1)} & \colhead{(2)} & \colhead{(3)} & \colhead{(4)} & 
\colhead{(5)} & \colhead{(6)} & \colhead{(7)} & \colhead{(8)} 
} 
\tablecolumns{8}
\startdata
NGC 232 &  0.21 & 0.24  & 0.14  & 0.13 & 0.15 & 0.40  & $-$2.62 \nl
NGC 1204 & 0.22 & 0.25 & 0.14  & 0.14 & 0.17  & 0.52  & $-$2.22 \nl
ESO 550-IG025 N  &  0.18 & \nodata & 0.08 & 0.18 & 0.11 & 0.35 & $-$2.51 \nl
IRAS 10057-3343  &  \nodata & \nodata & \nodata & \nodata & \nodata & \nodata &
$-$1.90 \nl
ESO 500-G034  &  0.18 & 0.18 & 0.11 & 0.17 & 0.15 & 1.36 & $-$1.90 \nl
ESO 436-G026  &  0.16 & 0.18 & 0.14 & 0.17 & 0.17 & 1.29 & $-$1.85 \nl
MCG+00-29-023 & \nodata & \nodata & \nodata & \nodata & \nodata & \nodata & \nodata \nl
Mrk 52  &  0.21 & 0.24 & 0.15 & 0.22 & 0.18 & 3.82 & $-$2.33 \nl
MCG-02-33-098 W  &  0.11 & \nodata & 0.07 & 0.10 & 0.05 & 2.13 & $-$2.15 \nl
Mrk 1344  &  0.20 & 0.23 & 0.14 & 0.15 & 0.16 & 0.70 & $-$2.17 \nl
ESO 527-IG07  &  0.20 & 0.21 & 0.13 & 0.17 & 0.16 & \nodata & $-$2.05 \nl
ESO 602-G025 &  0.12 & \nodata & 0.08 & 0.12 & 0.07 & 0.22 & $-$2.36 \nl
\cutinhead{Starbursts} 
Mrk 529   &  0.22 & 0.23 & 0.13 & 0.12 & 0.17 & 2.23 & $-$2.17 \nl
\cutinhead{AGN}
Mrk 955  &  0.19 & 0.21 & 0.12  & 0.16  & 0.16  & \nodata & $-$3.15 \nl
Mrk 1388  &  0.01 & 0.01 & 0.02 & 0.06 & 0.0 & 0.50 & $-$0.58 \nl
IRAS 16164-0746  &  0.24 & \nodata & 0.15 & 0.15 & 0.20 & 1.56 & $-$1.82 \nl
NGC 7130   &  0.16 & 0.18 & 0.11 & 0.12 & 0.13 & 0.46 & $-$2.50 \nl
\enddata
\tablecomments{CO$_{\rm{sp}}$ has been measured directly using 
equation~1 and a variety of wavelength regions. \nl
(2): wavelength region used by Ridgway {\it et~al.} (1994) \nl
(3): wavelength region used by Doyon {\it et~al.} (1994)  \nl
(4): extended region, 2.2931 -- 2.3200 $\mu$m,
used by Puxley {\it et~al.} (1997) \nl
(5): narrow region, 2.2931 -- 2.2983 $\mu$m, 
used by Puxley {\it et~al.} (1997)\nl
(6): region used by Goldader {\it et~al.} (1995) \nl
(8): powerlaw index, obtained by fitting the continuum to the 
form $f_\lambda \propto \lambda^{\beta}$ }
\end{deluxetable}


\begin{deluxetable} {lccccc}  
\small
\tablewidth{0pt}
\tablenum{7}
\tablecaption{Rest equivalent widths of absorption lines \label{ew}}
\tablehead{
\colhead{Name} & \colhead{H$\beta$} & \colhead{Na~I} & 
\colhead{Ca~I} 
& \colhead{CO} & \colhead{CO} \nl
& \colhead{$\lambda$0.4861} & \colhead{$\lambda$2.206} & \colhead{$\lambda$2.263} & 
\colhead{narrow} & \colhead{extended} \nl
\colhead{(1)} & 
\colhead{(2)} & 
\colhead{(3)} & 
\colhead{(4)} & 
\colhead{(5)} & 
\colhead{(6)}} 
\tablecolumns{6}
\startdata
NGC 232 & 0.33 & 1.59 & 2.86 & 0.60 & 3.15 \nl
NGC 1204 & 0.34 & 2.93 & 1.53 & 0.63 & 3.28 \nl
ESO 550-IG025 N & 0.63 & \nodata & \nodata & 0.78 & 1.83 \nl
IRAS 10057-3343 & \nodata & \nodata & \nodata & \nodata & \nodata \nl
ESO 500-G034 & \nodata & 2.24 & 1.49 & 0.73 & 2.57 \nl
ESO 436-G026 & \nodata & 5.62 & 3.88 & 0.77 & 3.15 \nl
MCG+00-29-023 & 0.31 & \nodata & \nodata & \nodata & \nodata  \nl
Mrk 52 & \nodata & \nodata & 2.45 & 0.97 & 3.52 \nl
MCG-02-33-098 W & 0.53 & \nodata & \nodata & 0.48 & 1.68 \nl
Mrk 1344 & \nodata & 3.79 & 2.32 & 0.69 & 3.23 \nl
ESO 527-IG07 & \nodata & \nodata & 1.71 & 0.75 & 2.93 \nl
ESO 602-G025 & 0.26 & \nodata & \nodata & 0.54 & 1.92 \nl
\cutinhead{Starbursts}
Mrk 529 & & \nodata & \nodata & 0.53 & 3.08 \nl
\cutinhead{AGN}
Mrk 955 & \nodata & 1.94 & \nodata & 0.72 & 2.74 \nl
Mrk 1388 & \nodata & \nodata & 1.51 & 0.27 & 0.60  \nl
IRAS 16164-0746 & 0.37 & \nodata & 3.05 & 0.68 & 3.38 \nl
NGC 7130 & \nodata & 2.36 & 2.37 & 0.56 & 2.68 \nl
\enddata
\tablecomments{All equivalent widths are given in nm. \nl
(2): H$\beta$ absorption taken from Kim {\it et~al.} (1995) \nl
(5): narrow range, 2.2931 -- 2.2983 $\mu$m, following Puxley {\it et~al.} 
(1997) \nl
(6): extended range, 2.2931 -- 2.3200 $\mu$m, following 
Puxley {\it et~al.} (1997)}
\end{deluxetable}

\end{document}